%% file: main.tex
\documentclass[11pt]{article}
\usepackage{setspace} 
\usepackage[utf8]{inputenc}
\usepackage[english]{babel}
\usepackage{graphicx}
\usepackage{caption}
\usepackage{subcaption}
\usepackage{ragged2e}
\usepackage{lscape}
\usepackage[colorlinks=true,allcolors=black]{hyperref}
\usepackage{multirow}
\usepackage{amsmath}
\usepackage[margin=2.5cm]{geometry}
\usepackage[capposition=bottom]{floatrow}
\usepackage{comment} 

\newcommand{\emil}[1]{}
\newcommand{\andreas}[1]{}
\usepackage[
authordate,bibencoding=auto,strict,backend=biber,natbib
]{biblatex-chicago}
\AtEveryBibitem{\clearfield{issn}}
\AtEveryBibitem{\clearfield{doi}}
\AtEveryBibitem{\clearfield{url}}
\AtEveryBibitem{\clearfield{month}}
\newtheorem{hypothesis}{Hypothesis}

\newcommand{\sym}[1]{{#1}}

\title{Why Do Students Lie and Should We Worry? An Analysis of Non-truthful Reporting
\footnote{We are thankful to Julien Grenet, Vincent Meisner, Li Chen, Jeppe Søndergaard Johansen, Jeanet Sinding Bentzen, Isabel Skak Olufsen, and peer feedback at Bristol School Choice Workshop, European Meeting of Econometrics 2021, Chinese Meeting of Econometrics 2021, Educational Workshop at Lund University 2021, and lunch seminars at the University of Copenhagen. In addition, we are grateful to Anne Toft Hansen and Helene Willadsen for sharing their survey design with us.}
}
\author{Emil Chrisander\footnote{University of Copenhagen, email: emil.chrisander@econ.ku.dk} \and Andreas Bjerre-Nielsen\footnote{University of Copenhagen, email: anbn@econ.ku.dk}}
\date{This version: \today}

\begin{document}
\maketitle

\begin{abstract}
A core aspect of market design is to encourage participants to truthfully report their preferences to ensure efficiency and fairness. 
Our research paper analyzes the factors that contribute to and the consequences of students reporting non-truthfully in admissions applications. We survey college applicants in Denmark about their perceptions of the admission process and personality to examine recent theories of misreporting preferences. Our analysis reveals that omissions in reports are largely driven by students' pessimistic beliefs about their chances of admission. Moreover, such erroneous beliefs largely account for whether an omission led to a missed opportunity for admission. However, the low frequency of these errors suggests that most non-truthful reports are "white lies" with minimal negative impact. We find a novel role of personality and individual circumstances that co-determine the extent of omissions. We also find that estimates of students' demand are biased if it is assumed that students report truthfully, and demonstrate that this bias can be reduced by making a less restrictive assumption. Our results have implications for the modeling of preferences, information acquisition, and subjective admission beliefs in strategy-proof mechanisms.
\end{abstract}

\section{Introduction}

\input{paper_sections/1_introduction}

\section{Hypotheses and Methodological Approach}\label{sec:hypotheses_methods}
\input{paper_sections/2_hypotheses_method_approach}

\section{Institutional Setting \& Data Summary}\label{sec:institution}
\input{paper_sections/3_institutional_settings_data}

\section{Why Do Students Report Non-truthfully?}\label{sec:why_misreport}
\input{paper_sections/4_non_truthful_reporting_analysis}

\section{What Causes Payoff-relevant Omissions?}\label{sec:welfare_misreport}
\input{paper_sections/5_payoff_relevant_mistakes_analysis}
\section{How Does Non-truthful Reporting Affect Demand Estimation?}\label{sec:demand_estimate_misreport}
\input{paper_sections/6_demand_estimation_analysis}
\section{Robustness Analyses}\label{sec:robust_analyses}
\input{paper_sections/robustness_analyses}

\section{Literature Review}\label{sec:lit_review}
\input{paper_sections/literature_review}
\section{Discussion}\label{sec:discussion_and_conclusion}
\input{paper_sections/discussion_and_conclusion}

\newpage
\printbibliography[]

\section{Appendix}
\input{paper_sections/X_appendix}

\end{document}

%% file: paper_sections/1_introduction.tex

An expanding literature investigates how to design matching mechanisms, in particular the school choice problem of how to allocate students among preferred schools and colleges \citep{Abdulkadiroglu2003SchoolApproach}. A key goal of these mechanisms is to encourage students to report their true preferences honestly, and strategy-proof mechanisms achieve this by ensuring that students cannot benefit by misrepresenting their preferences. However, recent evidence from laboratory experiments and field studies shows that a substantial number of students report non-truthfully, even in the presence of a strategy-proof mechanism  \citep[see, e.g.,][]{hassidim2021limits}. 
This raises important policy questions about the potential harm to students from such behavior, the underlying factors that contribute to it, and the consequences of non-truthful reporting for measuring students' demand.

Despite the significance of this issue, previous field studies have largely been limited to descriptive analysis or structural modeling due to a lack of access to data on students' backgrounds and perceptions. Our research paper fills this gap by examining the determinants and consequences of non-truthful reporting behavior in the context of higher education admissions in Denmark, where the mechanism is strategy-proof. We surveyed the universe of higher education applicants in Denmark over the years 2020 and 2021 about their admission perceptions, and combined this data with registry data and administrative records of student choices to analyze non-truthful behavior.\footnote{The admission is based on DA with Voluntary Information Disclosure, which allows for endogenous application to a subset of seats with additional screening \citep{Bjerre-Nielsen2022VoluntaryProperties}. We also regard the issue of constrained school choice \citep{Calsamiglia2010ConstrainedStudy} in our context as a negligible problem. This is due to the fact that only 2\% of applicants use the maximum of eight available options and that we find those who use them all have similar non-truthful reporting rates to those who do not, see Institutional Setting \& Data Summary.} 

We document that many students do not report truthfully in  the Danish context and that this behavior leads to missed admission opportunities. We depart from previous work by estimating models of the decision to report truthfully using detailed administrative and socio-demographic data, as well as self-reported perception and personality data. Our findings indicate that students' subjective beliefs about their chances of admission are a major contributor to non-truthful behavior, and that students' perceptions and personalities, as well as their willingness to postpone enrollment, also play a role. In terms of consequences, we find that errors in subjective beliefs are responsible for the majority of missed admission opportunities, highlighting the importance of admission transparency. Finally, we show that estimates of students' preferences are substantially biased if it is assumed that students report truthfully, and demonstrate that this bias can be reduced by making a less restrictive assumption.




We begin our analysis by presenting a descriptive analysis of applicants' non-truthful behavior and quantify some of this behavior leads to missed admission opportunities, which is consistent with existing studies. Among our survey respondents, 20\% report non-truthfully in the sense that they agree with the statement that they would change their rank-ordered list if they could be admitted to any education program. The majority of these (60\%)  omitted their most-preferred option. To assess whether the students would have been admitted by reporting truthfully we compare the realized cutoffs with students' own eligibility scores. Among the students who report non-truthfully, 10\% would have obtained an improved match had the students reported truthfully.

To understand such strategic behavior, we estimate models of the decision to report truthfully using exact administrative and socio-demographic data along with self-reported perception and personality. 
Using individual measures of admission beliefs for programs applied to or deliberately omitted allows us to analyze the role of both rational and subjective beliefs. 
Our estimates show that going from believing admission is certain (100\% chance of admission) to impossible (0\% chance) substantially increases the likelihood of reporting non-truthfully by 31 percentage points.\footnote{When we include program level fixed effects the effect is 33 pct. points.} These estimates increase substantially when we incorporate the beliefs about entering the most preferred program through an alternative admission criterion;\footnote{See Institutional Setting \& Data Summary for a description of the admission system, including the main and alternative admission criteria and quotas.} when the combined belief goes from certain to impossible the non-truthful reporting increases by 40 percentage points. 
Thus, admission beliefs alone account for a large share of the variation in non-truthful behavior. 

As a novelty, we analyze what causes non-truthful behavior to have payoff consequences. We decompose subjective beliefs into objective beliefs and belief errors \citep{Agarwal2018DemandMechanism}. We find that subjective belief errors play a further decisive role in accounting for the welfare loss from non-truthful reporting. Specifically, a non-truthful student with pessimistic admission beliefs is 54\% more likely to incur a welfare loss. 

To investigate other determinants of non-truthful reporting by students we collect data on their personality along with other perceptions of the admission system and combine the measures with registry data for socio-demographic background and prior academic achievements. We show that measures of self-confidence, risk-willingness and perception of the mechanism's transparency, in particular the concept strategy-proofness, increase the likelihood of truthful reporting. In addition, outside options captured by a student's  willingness to postpone enrollment by an additional year also increase truthful reporting.
Finally, we show that academic ability and socio-demographic factors, parents' education and income, and own sex, do not account for non-truthful behavior.

An additional concern of non-truthful reporting is that it can have severe consequences for demand estimation of educational preferences \citep{Fack2019BeyondAdmissions, Larroucau2019DoProblem}. Specifically, the critical assumption of weak-truthtelling implies that no first choices are omitted, is used in the majority of state-of-the-art econometric methods and applied work \citep[see, e.g.,][]{Kapor2020HeterogeneousMechanisms,agarwal2020revealed}.
We assess this assumption by analyzing whether accounting for non-truthful measures affects the demand estimation. To accomplish this, we compare three different approaches: (i) a naive approach that applies students' reported first choice first; (ii) a revealed-preferences approach that applies the stated first choice from our survey; (iii) a stability-based approach, which is currently one of the state-of-the-art approaches \citep{Fack2019BeyondAdmissions,Artemov2021StrategicResearch,Arslan2021PreferenceLists}.\footnote{See the literature review for a more detailed account of relevant estimation methods.} The latter approach applies students' reported first choices against all other feasible programs where feasibility is determined by admission cutoffs \citep{Fack2019BeyondAdmissions}. We find that the naive approach yields estimates that are severely biased compared to the revealed-preferences. Yet, we find that the stability-based approach yields similar results to the elicitation-based approach, which serves as further validation of the stability-based approach.

Our findings contribute to the literature on non-truthful reporting in three distinct ways. First, it provides evidence from the field for theories that use subjective beliefs about admission to accounting for the non-truthful reporting, e.g., ego-utility \citep{Koszegi2006EgoChoice}, expectations-based loss aversion \citep{Dreyfuss2022Expectations-BasedMechanisms} and report-dependent utility \citep{Meisner2022Report-DependentStrategy-Proofness}. These theories account for the avoidance of rejection using subjective beliefs about admission chances. Consistent with these theories, \citet{Li2017ObviouslyMechanisms,Dreyfuss2022Expectations-BasedMechanisms} show, in economic experiments, that when a subject is endowed with a lower priority score, and thus faces a higher likelihood of rejection at the most popular schools, they are more likely to misrepresent their preferences.  Second, we show a clear role for individual circumstances and personal characteristics, which contributes to the understanding of why strategy-proofness fails to materialize in real-world settings. In particular, we find a novel role for economic and psychological characteristics. 
Conversely, our findings that socio-demographic background plays a negligible role is contrary to \citet{Chen2019Self-selectionChoice}.\footnote{This is likely because they do not have data on admission beliefs, which reduces the coefficients on socioeconomic background in our setting.} Finally, we provide further evidence using our novel data against using the weak-truth telling assumption in preference estimation and we validate the stability-based approach to estimating preferences as an alternative. This complements and supports the existing validation of the stability-based approach by structural models in \citet{Fack2019BeyondAdmissions}.

Two essential policy implications emerge from our findings. First, the novel role of students' perception and personality for non-truthful reporting and its consequences demonstrates the welfare potential for information provision. A related scope has already been demonstrated using admission advice focused on students at risk of being rejected at all places they apply to \citep{Arteaga2022SmartChoice}. Second, our result that non-truthful behavior is critical for credibly measuring student preferences underscores the importance of adjusting for omitted programs. 
Nevertheless, we conclude that while we can say Danish students lie about their preferences, these lies are predominantly "white lies" as they usually are inconsequential for their own or others' admission prospects.

Our findings are likely to have external validity and be applicable in other contexts of student admission worldwide. We draw this conclusion from the fact that there is a high similarity between our descriptive evidence of non-truthful behavior and its consequences and that of existing field studies across multiple continents, see Literature Review. In an international comparison, the Danish admission system to higher education is characterized by a high degree of transparency and admission predictability.\footnote{Note that the subset of admissions based on the alternative admission criterion are less transparent as cutoffs are not public and the admission criteria are also less clear, see Institutional Setting \& Data Summary for details about the system. Our results are, however, robust to extending beliefs to cover admission in this manner.} Thus, a consequence of our finding that belief errors lead to a payoff-relevant omission is that less transparent admission systems will see more missed admission opportunities caused by non-truthful reporting - this fact is consistent with us measuring fewer payoff-relevant mistakes than \citet{Chen2019Self-selectionChoice}.


The remainder of this paper is organized as follows. Initially, we present our hypotheses and methodological approach. Hereafter, we introduce the institutional setting, data sources and descriptive statistics in Section~\ref{sec:institution}. Our analysis is divided into three parts: what causes non-truthful reporting in Section~\ref{sec:why_misreport}; what are the welfare consequences of non-truthful reporting in Section~\ref{sec:welfare_misreport}; finally,  Section~\ref{sec:demand_estimate_misreport} analyzes the consequence of non-truthful reporting in demand estimation. Section~\ref{sec:robust_analyses} contains robustness analyses. Section \ref{sec:lit_review} reviews the relevant literature and compares with our findings. Section~\ref{sec:discussion_and_conclusion} discusses our results and conclusions. 
The Appendix contains auxiliary results and explanations.

%% file: paper_sections/2_hypotheses_method_approach.tex
In this section, we outline our outcome variables, hypotheses, and the methods we will use to test them. We refer to the origin of the hypotheses as some are motivated by economic theory while others by lab experiments.




\paragraph{Definition of outcome variables} The main outcome variables of our paper are \textit{non-truthful reporting}, \textit{omitting most-preferred program}, and \textit{making a payoff-relevant omission}. We define a survey respondent as reporting non-truthfully if she answers "yes" to the following question: "Would you change your submission if you could be admitted to any study program?" We define a student as omitting her most-preferred program if she reports non-truthfully and answers "yes" to the following question: "Would you change your top-ranked choice if you could be admitted to any study program?" Henceforth, we define a student's revealed most-preferred program based on her truthful answer. If she reports truthfully, we define her revealed most-preferred program as the top choice in her ROL. If she does not report truthfully, we ask the student to state her most-preferred program in the survey and use this answer as her revealed most-preferred program. Finally, we define a respondent as making a payoff-relevant omission if she would have been admitted to her omitted most-preferred program.

\paragraph{Determinants of non-truthful behavior}
We begin with the theory that students omit prominent programs when they perceive admission to be unlikely or impossible. We operationalize this in Hypothesis~\ref{hypothesis:beliefs}.

\begin{hypothesis}
\label{hypothesis:beliefs} Students report non-truthfully and omit most-preferred programs if they believe the probability of being admitted into their most-preferred program is zero or very low.  
\end{hypothesis}

We test Hypothesis~\ref{hypothesis:beliefs} in the following model:
\begin{align}
    y_i = \alpha + \beta X_{H1}+\Theta+\epsilon_i
    \label{eq:model_hyp_1}
\end{align}

\noindent Where $y_i$ is the outcome of interest, and $X_{H1}$ includes students' stated admission beliefs for their most-preferred program. $\Theta$ is an indicator of survey wave. We proceed by presenting hypotheses that consider determinants of payoff-relevant omissions. To measure the relative importance of admission beliefs, we introduce two additional hypotheses for non-truthful reporting. The second hypothesis is motivated by the fact that existing work have found an explanatory role of prior academic achievements and cognitive abilities as well as socioeconomic for non-truthful behavior background \citep{Hassidim2016StrategicEnvironment,Chen2019Self-selectionChoice}. Unlike the existing studies, we use registry data to exactly measure the information. We measure student SES as their parents' income rank within their cohort and years of education.\footnote{We consider the cohort as the sub-population of same gender and age as the parent.}

\begin{hypothesis}
\label{hypothesis:SES} Students' demographics, socioeconomic status, and academic achievements affect their propensity to report non-truthfully and omit most-preferred programs.
\end{hypothesis}

The third hypothesis is motivated by a number of previous findings. \citet{Dreyfuss2022Expectations-BasedMechanisms,Meisner2022Report-DependentStrategy-Proofness} show theoretically that dis-utility from rejection may affect truthful reporting. We include a number of measures from the survey, see the next section for an overview and an account of why they are included.

\begin{hypothesis}
\label{hypothesis:Personality} Students' personalities and life situations affect their propensity to report non-truthfully and omit most-preferred programs.
\end{hypothesis}

We test the Hypotheses \ref{hypothesis:beliefs}-\ref{hypothesis:Personality} simultaneously in the following model:

\begin{align}
    y_i = \alpha + \beta X_{H1}+\Gamma X_{H2}+\Delta X_{H3}+\Theta+\epsilon_i
    \label{eq:model_hyp_1_2_3}
\end{align}

\noindent Where $y_i$ is the outcome of interest, and $X_{H1}$ includes students' stated admission beliefs for their most-preferred program\footnote{We apply a continuous variable since the conclusion of our results remains the same as when it is coded as a discrete variable}. The vector of variables, $X_{H2}$, includes the socio-demographic information and prior academic achievements for the second hypothesis. $X_{H3}$ contains the self-reported measures for the third hypothesis. Finally, $\Theta$ is an indicator of survey wave. We proceed by presenting hypotheses that consider determinants of payoff-relevant omissions.

\paragraph{Determinants of payoff-relevant omission} Once more we consider the theory that students omit prominent programs when they perceive admission to be unlikely or impossible. This theory suggests that payoff-relevant omissions can occur if a student's subjective admission beliefs are pessimistic compared to rational admission beliefs. We operationalize this in Hypothesis ~\ref{hypothesis_mistake:beliefs_errors}.

\begin{hypothesis}
\label{hypothesis_mistake:beliefs_errors} Payoff-relevant omissions of most-preferred programs are caused by students' pessimistic admission beliefs. 
\end{hypothesis}

To measure the relative importance of pessimistic admission beliefs, we introduce two additional hypotheses for non-truthful reporting. The motivation for these is the same as for Hypotheses~\ref{hypothesis:SES} and ~\ref{hypothesis:Personality}.

\begin{hypothesis}
\label{hypothesis_mistake:SES} Students' demographics, SES, and academic achievements affect their propensity for payoff-relevant omissions of their most-preferred programs.
\end{hypothesis}

\begin{hypothesis}
\label{hypothesis_mistake:Personality} Students' personalities and life situations affect their propensity for payoff-relevant omissions of their most-preferred programs.
\end{hypothesis}

We test the Hypotheses \ref{hypothesis_mistake:beliefs_errors}-\ref{hypothesis_mistake:Personality} simultaneously in the following model:
\begin{align}
    y_i = \alpha + \beta X_{H4}+\Gamma X_{H5}+\Delta X_{H6}+\Theta+\epsilon_i
    \label{eq:model_hyp_4_5_6}
\end{align}

\noindent where $y_i$ is whether an omitted most-preferred program is a payoff-relevant omission. $X_{H4}$ includes indicator variables of whether a student is pessimistic or optimistic. $X_{H5}$ includes the socio-demographic information and prior academic achievements. $X_{H6}$ contains the self-reported measures. Finally, $\Theta$ is an indicator of survey wave. We proceed to highlight the institutional setting and present descriptive statistics. 


%% file: paper_sections/3_institutional_settings_data.tex
Our analysis focuses on the annual Danish higher education central admission system. It is a Deferred Acceptance-like mechanism with approximately 90,000 students and 900 study programs. Students can apply to a maximum of eight study programs. These can be academic and non-academic study programs. In this paper, we consider a subgroup of students who apply to at least one academic study program. The mechanism applies an eligibility score as the primary method used to determine which students to admit into a study program. The eligibility score is based on students' high school GPA. The Danish admission system allows study programs to admit a minor share of students through alternative evaluation criteria\footnote{Every study program set their own evaluation criteria. Common admission criteria are specific high-school grades, submitted essays, multiple-choice tests, and interviews.} than eligibility score. \citet{Bjerre-Nielsen2022VoluntaryProperties} prove that this type of mechanism is ordinal strategy-proof.\footnote{This implies that submitting a truthful report according to ordinal preferences remains a weakly dominant strategy.} Nonetheless, as pointed out by \citet{Haeringer2009ConstrainedChoice} students face a strategic incentive if they consider more programs than they can apply to in a ROL. However, in the Danish context 98\% of the applications contains less programs than the limit. In addition, among students who submit a ROL with 8 programs there is no significant difference in the non-truthful rate compared to students who submit less than 8 programs. As a result, we do not expect a constrained ROL to affect the conclusion of our findings.    

\begin{figure}[!ht]
    \begin{subfigure}[b]{.495\linewidth}
        \includegraphics[width=\linewidth]{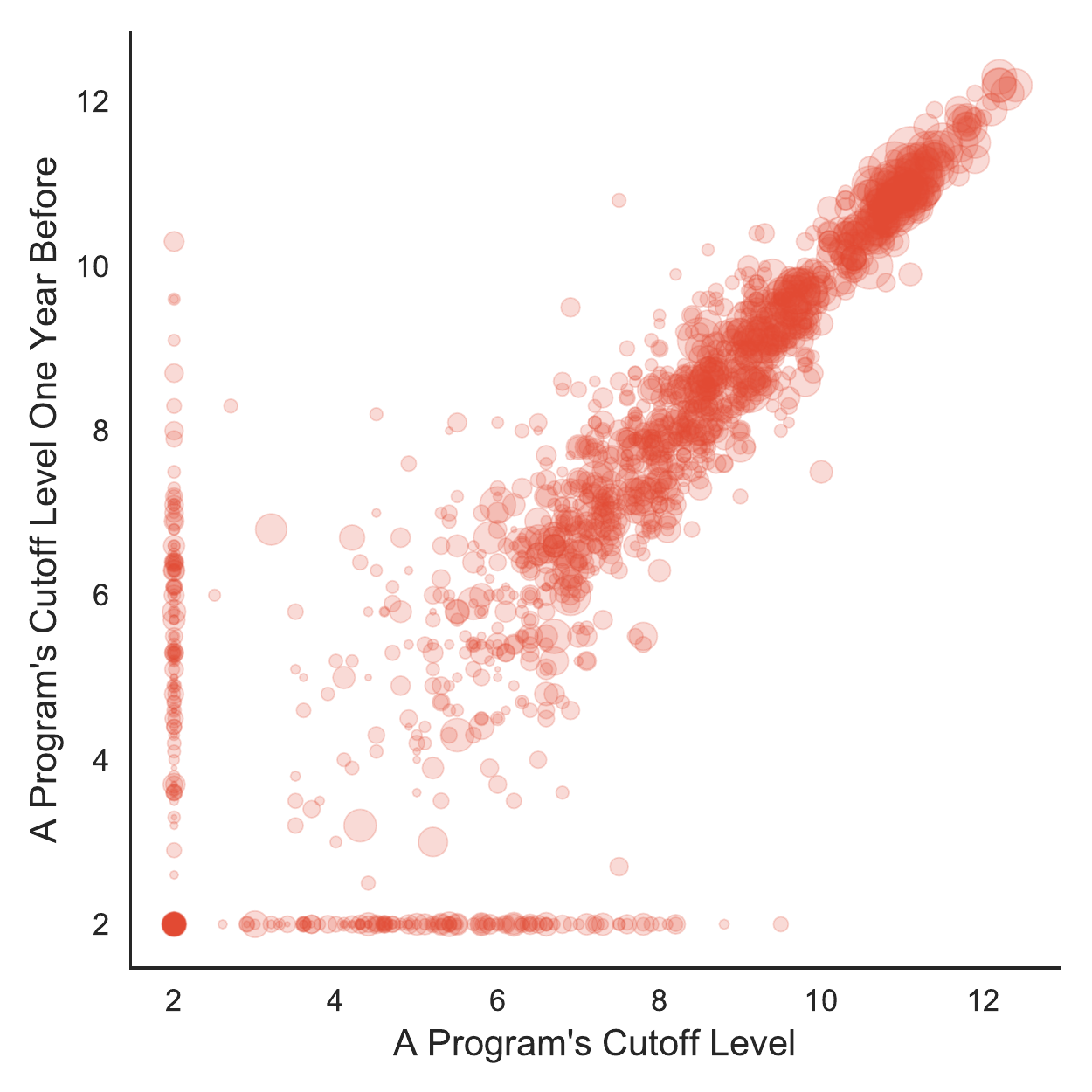}
        \caption{Stability of Program Cutoff Levels Across Time}
        \label{fig:scatter_cutoff}
    \end{subfigure}
    \begin{subfigure}[b]{.495\linewidth}
        \includegraphics[width=\linewidth]{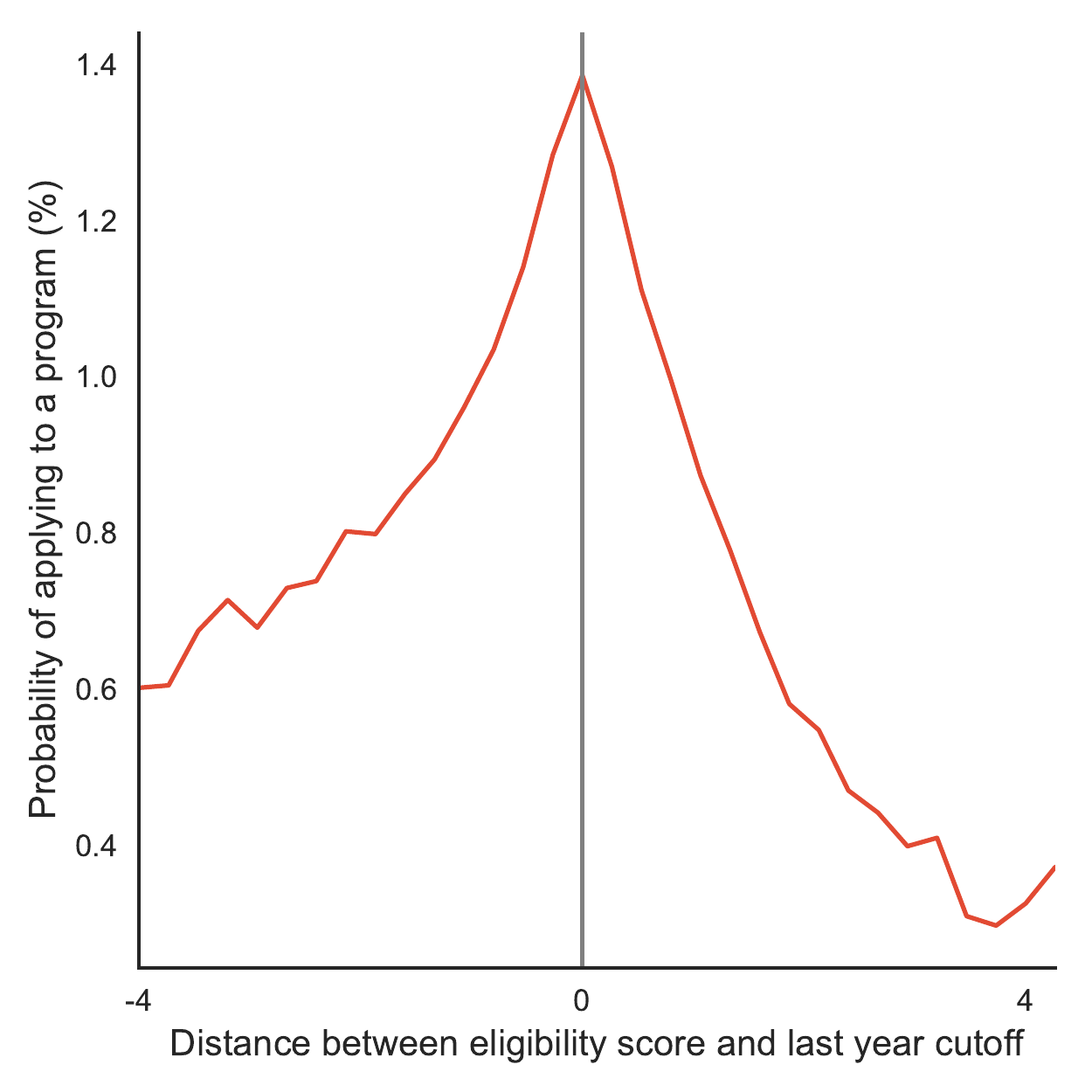}
        \caption{Relative eligibility score and application likelihood}
        \label{fig:apl_prob_dist_cutoff}
    \end{subfigure}
\caption{Key Insights From The Central Danish Admission System}
\label{fig:stylized_facts}
\floatfoot{Notes: The left-hand figure is a scatter plot with a program's cutoff level on the x-axis and the same program's cutoff level from one year before on the y-axis. The size of the marker indicates the number of applicants to a program, where a larger (smaller) size represents a larger (smaller) program. The right-hand figure plots the probability of applying to a program conditional on the distance between eligibility score in 2019 and the cutoff level in 2018. The probability of applying to a program is calculated as the average share of applicants who apply to a program within bins of distance to eligibility score. A total of 33 bins is used to generate the plot. Only programs with a cutoff level above 9 are considered to ensure sufficient mass around the cutoff.}
\end{figure}

To make our results comparable to mechanisms in other countries, we omit applications to study programs using the alternative criteria. Similar approaches of using an eligibility score\footnote{Typically the eligibility score is based on previous academic achievements or a standardized test score} to prioritize students are applied in strategy-proof school choice mechanisms in a number of developed and developing countries.\footnote{The list of countries include Australia, Chile, Finland, Ghana, Hungary, Ireland, Kenya, Lesotho, Liberia, Mexico, Norway,  Romania, Spain, Singapore, Trinidad and Tobago, Turkey, Tunesia, Uganda, USA, and Zambia \citep{Ajayi2020WhenSystem,Fack2019BeyondAdmissions}.} These mechanisms produce cutoff levels that denote the minimum eligibility score required for admission into a study program. Thus, the probability of admission into a program for a student corresponds to the probability of a student's eligibility score being equal to or higher than the cutoff score for the program. In the Danish context, students know their eligibility score at the time of application. Moreover, every year the post admission cutoff levels are announced and receive much public attention. Finally, as shown in Figure~\ref{fig:scatter_cutoff}, the study program cutoffs levels are relatively stable across time. As a result, students can use historic cutoff levels to form beliefs about forthcoming cutoff levels, and thereby, their admission probabilities. 

We highlight the historic cutoff levels because they seem to play a decisive role in students' admission behavior. Specifically, students tend to apply to study programs with historic cutoff levels very close to their own eligibility score.\footnote{The same behavioral pattern is observed in Chile \citep{Larroucau2019DoProblem} and Turkey \citep{Arslan2021PreferenceLists}} To illustrate this, Figure~\ref{fig:apl_prob_dist_cutoff} plots the probability of applying to a study program conditional on how close a student's eligibility score is to the previous year's published cutoff level. It is puzzling that there is such a strong kink in the slope around the previous year's cutoff level and that students are less likely to prefer very selective programs unless they have a sufficiently high eligibility score. We are surprised to find this type of admission behavior in a strategy-proof admission system, because students cannot be worse of by applying to selective programs in a strategy-proof mechanism. Although Figure~\ref{fig:apl_prob_dist_cutoff} indicates that students do not report truthfully, it cannot rule out other explanations that are consistent with truthful reporting. E.g., if students care about the peer composition of programs, it may be the case that students have stronger preferences for programs with cutoff levels close to their own eligibility score. This problem motivates our survey approach to identify non-truthful reporting.

\subsection{Data sources and measures}

We now outline our three sources of data. The first source of data is application and admission data from the centralized higher education admission system in Denmark. Importantly, the data contain students' reported preferences for study programs, i.e., their rank-ordered list (ROL). We use data from 2020 and 2021 for our analysis. The number of students applying is relatively stable across the years, although an unusually high number of students applied in 2020. The number of academic study programs has decreased since 2012 and ranges from around 320-350. About 50\% of the study programs receive more applications than their capacity, resulting in a cutoff level for those study programs. We refer to table \ref{tab:admission_summary_stats} in the appendix for further statistics on the admission system.

Our second source of data is Danish registry data on applicants and their parents. This data contains socioeconomic and demographic variables. A significant advantage of registry data is the fact that it covers the entire population of
Denmark, and it is maintained by Statistics Denmark for research purposes.

Finally, we conducted two waves of an online survey targeting students who applied to higher education in Denmark. The first wave was in 2020 and the second wave was in 2021. The timing of the surveys was such that students knew their eligibility score, but had no knowledge about the realized cutoff levels.\footnote{The students received an invitation to complete the survey three days after the application deadline. The survey was open for responses until the day before the students received a personal matching answer. The cutoff levels of every study program are announced on the same day students receive a matching answer. Thus, none of the survey respondents knew the realized cutoff levels at the time of answering the survey. However, the students did know their final eligibility score at the time of answering the survey.} The survey response rate was $15\%$. Table~\ref{tab:pop_vs_survey} compares the mean of indicator variables for the population and survey respondents. 

In addition, we include a column for the sample of survey respondents included in the regression analyses. The regression sample includes survey respondents without any missing data.

\input{tables/pop_vs_survey_2021}

\paragraph{Explanatory variables}
From the registry, we include measures of students' age at admission, a woman indicator, eligibility score, and middle school GPA. In addition, we include the income percentile and years of education for students' parents.

From our survey we elicit a number of additional measures of students perceptions and personal situation. We measure whether students `Perceive rejection as being a failure'. A similar concern is that low ability students are more likely to report non-truthfully and omit preferred programs \citep{Chen2019Self-selectionChoice}. We address this concern by measuring `Confidence in own abilities'. We also explore whether skipping is more common among students who do not fully comprehend the admission process, for instance if they misperceive the admission procedure to be Immediate Acceptance \citep{Abdulkadiroglu2003SchoolApproach}. Finally, students who do not mind postponing admission to next year are presumably less concerned with not being admitted to a study program \citep{akbarpour2022centralized}, which we investigate by measuring `Willingness to postpone'. See the exact definition of the variables and how they are measured in Appendix~\ref{app:survey_var_describe}.

\paragraph{Selective participation}
An inherent concern of survey data is whether the participants are representative of the population of interest. In Table \ref{tab:pop_vs_survey} we report a two-sided Welch's t-test to compare the statistical differences in registry variables between our population and survey participants. Compared to the population, the survey respondents are marginally younger, more likely to be women, and have a somewhat higher eligibility score and middle school GPA. Moreover, they tend to report marginally more choices in their ROL. In Section \ref{sec:robust_analyses}, we apply a Weighted Least Square approach suggested by \citep{Dutz2021SelectionSurveys} as a robustness check for selective participation. We conclude that selective participation does not affect our findings and conclusions.

\subsection{Descriptive analysis of survey data}
We show descriptive statistics of our outcome and explanatory variables for our sample (and population where possible) in Table~\ref{tab:pop_vs_survey}. In terms of outcomes, 20\% of the students report non-truthfully, 12\% omit their most-preferred program, and 2\% make a payoff-relevant omission.

Most of the explanatory variables measures are well-established in the literature. \footnote{e.g., sociodemography, academic achievements as well as confidence and economic risk preferences}. However, we elicited a number of new measures which have not been analyzed before.
First and foremost, a striking 55\% respond that not getting admitted is a failure. This provides direct support for theories that incorporate dis-utility from rejection, either directly from e.g. report-dependent preferences \citep{Meisner2022Report-DependentStrategy-Proofness} or indirectly from Expectation-Based Loss Aversion \citep{Dreyfuss2022Expectations-BasedMechanisms}. 
Moreover, 38\% respond that they find admission system difficult to comprehend. This highlights the potential for policies that alleviate this problem.
Finally, 70\% of respond are willing to postpone their admission by an additional year. This number is extremely large an indicates that viewing the admissions in a single year in isolation may miss out on crucial dynamic aspects \citep{larroucau2020dynamic}.

%% file: tables/pop_vs_survey_2021.tex
\begin{table}[!ht]
\centering\small
\begin{tabular}{l|ccc|c} \hline 
 & & \textbf{Sample} & & \textbf{Sample diff.}\\
\textbf{Category}          & \textit{Population} & \textit{Survey resp.} & \textit{Reg. Sample} & \textit{Pop. vs. Survey} \\ \hline
 \textit{Survey Waves} &                 &             &        &        \\
 Respondent in 2020 wave             & -         & 0.55         & 0.53  & -          \\
 Respondent in 2021 wave             & -         & 0.45         & 0.47  & -          \\
  &                 &             &          &      \\
\textit{Outcome Variables} &                 &             &        &        \\
Reports non-truthfully              & -         & 0.20         & 0.19  & -          \\
Omits most-preferred program              & -         & 0.12         & 0.11  & -          \\
Makes payoff-relevant omission       & -         & 0.02         & 0.02  & -          \\ 
 &                 &             &          &      \\
\textit{Survey Variables} &                &             &        &        \\
Confidence in own abilities                & -         & 6.72 (2.15)         & 6.69 (2.14)  & -          \\
Risk willingness                           & -         & 5.56 (2.17)         & 5.52 (2.13)  & -          \\
Willingness to postpone an additional year & -         & 0.70 (0.46)         & 0.71 (0.45)  & -          \\ 
Perceives rejection as failure             & -          & 0.55 (0.50)        & 0.56 (0.50)  & -          \\ 
Difficult to comprehend admission process  & -         & 0.38 (0.49)         & 0.38 (0.48)  & -          \\ 
 &                 &             &          &      \\
 \textit{Registry Variables} &                 &             &        &        \\
Age               & 22.0 (5.15)           & 21.7 (5.77)        & 20.6 (2.47) & -0.30\sym{***}           \\
Women             & 0.56 (0.50)            & 0.64 (0.48)        & 0.63 (0.48) & \textcolor{white}{-}0.08\sym{***}           \\
Eligibility score & 8.28 (2.05)             & 8.65 (2.03)       & 8.85 (1.97)  & \textcolor{white}{-}0.37\sym{***}           \\
Middle school GPA & 8.39 (1.92)             & 8.75 (1.88)       & 8.95 (1.79)  & \textcolor{white}{-}0.36\sym{***}           \\
Income percentile, parents  & 0.57 (0.26)            & 0.58 (0.25)        & 0.59 (0.23)& \textcolor{white}{-}0.01\sym{***}           \\
Years of education, parents  & 14.8 (2.80)           & 14.9 (2.64)       & 15.1 (2.51) & \textcolor{white}{-}0.10\sym{***}           \\
 &                 &             &          &      \\
\textit{Number of Choices} &                 &             &        &        \\
1                 & 0.40            & 0.36       & 0.35 & -0.04\sym{***}          \\
2                 & 0.23            & 0.23       & 0.24 & \textcolor{white}{-}0.00\sym{*}\textcolor{white}{\sym{**}}           \\
3                 & 0.15            & 0.17       & 0.17 & \textcolor{white}{-}0.02\sym{***}           \\
4                 & 0.10            & 0.11       & 0.11 & \textcolor{white}{-}0.01\sym{***}           \\
5                 & 0.05            & 0.06       & 0.05 & \textcolor{white}{-}0.01\textcolor{white}{\sym{***}}           \\
6                 & 0.03            & 0.03       & 0.04 & \textcolor{white}{-}0.00\sym{*}\textcolor{white}{\sym{**}}            \\
7                 & 0.02            & 0.02       & 0.02 & \textcolor{white}{-}0.00\textcolor{white}{\sym{***}}           \\
8                 & 0.02            & 0.02       & 0.02 & \textcolor{white}{-}0.00\sym{***}           \\ \hline
N                 & 85,882          & 13,352      & 8,049&                \\ \hline
\end{tabular}
\caption{Mean and standard deviation by sample}
\label{tab:pop_vs_survey}
\floatfoot{Notes: Means of indicator variables for sample universe and surveyed population. Parentheses contain standard deviation. 'Population' is every student applying to higher education in Denmark in 2020 and 2021. 'Survey' contains respondents who answered a population survey. Applicants received a survey invitation after the deadline of application to higher education. The survey was open for answers until the day before the announcement of cutoff levels. 'Reg Sample' includes the survey respondents included in the regression models. These are the survey respondents without missing data for covariates. 'Pop vs Survey' displays difference between population and survey mean. Please see the section 'Explanatory variables' for an explanation of the 'Survey Variables' and 'Registry Variables'. 'Number of Choices' presents count of study programs in students' submitted rank-ordered lists. P-values from a two-sided Welch's t-test is indicated by \sym{*} \(p<0.05\), \sym{**} \(p<0.01\), \sym{***} \(p<0.001\). }
\end{table}

%% file: paper_sections/4_non_truthful_reporting_analysis.tex
In this section, we analyze why students report non-truthfully and why they omit their most-preferred program. 

\begin{figure}[!ht]
    \begin{subfigure}[b]{1\linewidth}
        \includegraphics[width=\linewidth]{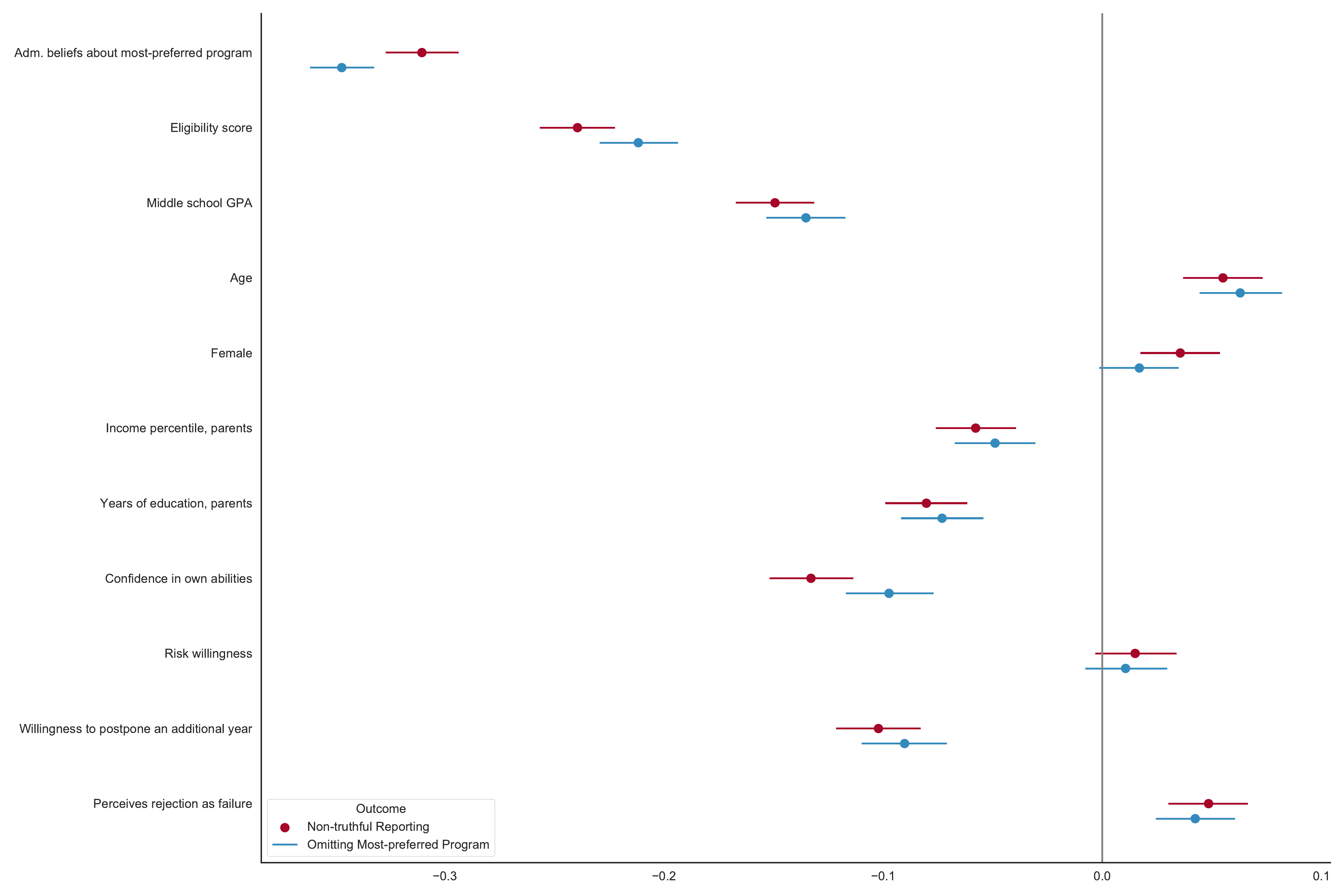}
    \end{subfigure}
\caption{Correlation between measures of truthfulness and explanatory variables}
\label{fig:correlation_non_truthful_explanatory}
\floatfoot{Notes: The dots mark the observed correlations, and the lines mark the 95\% confidence intervals. We calculated the confidence intervals by bootstrapping the data 10,000 times, calculating the pairwise correlations on each of the bootstrapped samples and taking the $5^{th}$ and $95^{th}$ percentile as the endpoints of the intervals.}
\end{figure}

We begin with examining the correlation between the explanatory variables and the measures of non-truthful behavior, see Figure \ref{fig:correlation_non_truthful_explanatory}. The figure shows that respondents' traits such as being older or female, or perceiving admission rejection as a failure all correlate with non-truthful behavior. Conversely, the following traits correlate to truthful behavior: i) high subjective admission beliefs, ii) high prior academic achievements (middle- or high-school grades), iii) high self-confidence, iv) high socioeconomic status (parental income or education), or v) willingness to postpone admission. 
These correlations point to a tendency of those being academic confidence and achievements or parents with high resources are more likely to exhibit truthful behavior. However, the above correlations may be due to variation in other observed and unobserved individual characteristics. To establish more reliable estimates we proceed with estimating the models outlined in Section~\ref{sec:hypotheses_methods}.

\paragraph{Admission beliefs and non-truthful behavior}

\begin{figure}[!ht]
    \begin{subfigure}[b]{1\linewidth}
        \includegraphics[width=\linewidth]{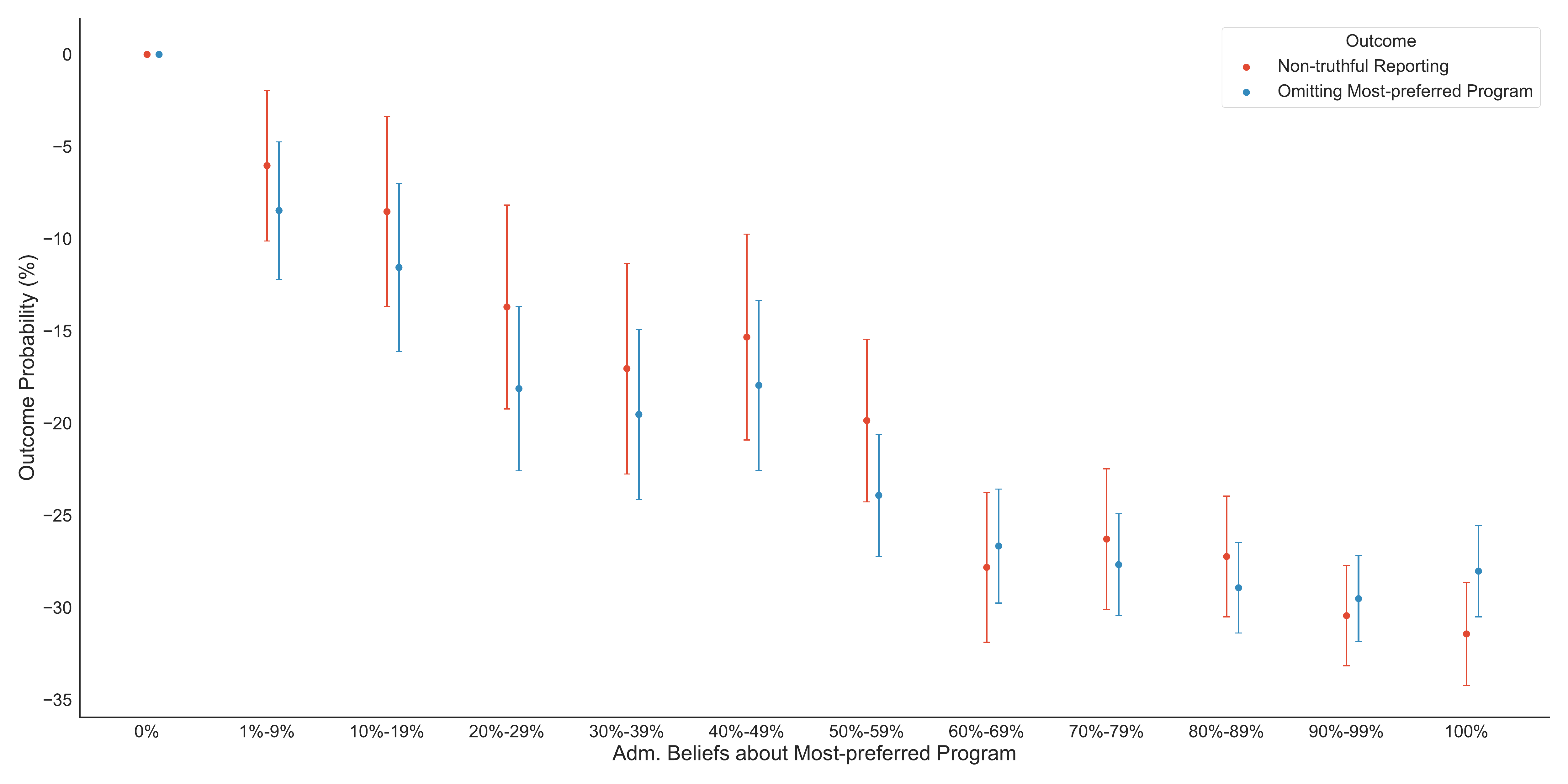}
    \end{subfigure}
\caption{Admission beliefs and truthful behavior; testing Hypothesis \ref{hypothesis:beliefs}}
\label{fig:hyp_1_ols_no_fe}
\floatfoot{Notes: The figure contains estimated coefficients from model (\ref{eq:model_hyp_1}) with 95\% confidence intervals. The dependent variable is one of two measures of non-truthful behavior, see the legend. The horizontal axis measures admission beliefs categories and the vertical axis the associated difference in the likelihood of non-truthful behavior. Standard errors of coefficients use robust estimation but are not clustered. }
\end{figure}

Figure \ref{fig:hyp_1_ols_no_fe} shows the estimation results of model \eqref{eq:model_hyp_1}. We observe that both non-truthful reporting and omission of most-preferred program decrease gradually as a student's admission beliefs increase. To be specific, the probability of non-truthful reporting decreases by 33 percentage points if a student's admission beliefs increase from 0\% to 100\%.

\begin{figure}[!ht]
    \begin{subfigure}[b]{.9\linewidth}
        \includegraphics[width=\linewidth]{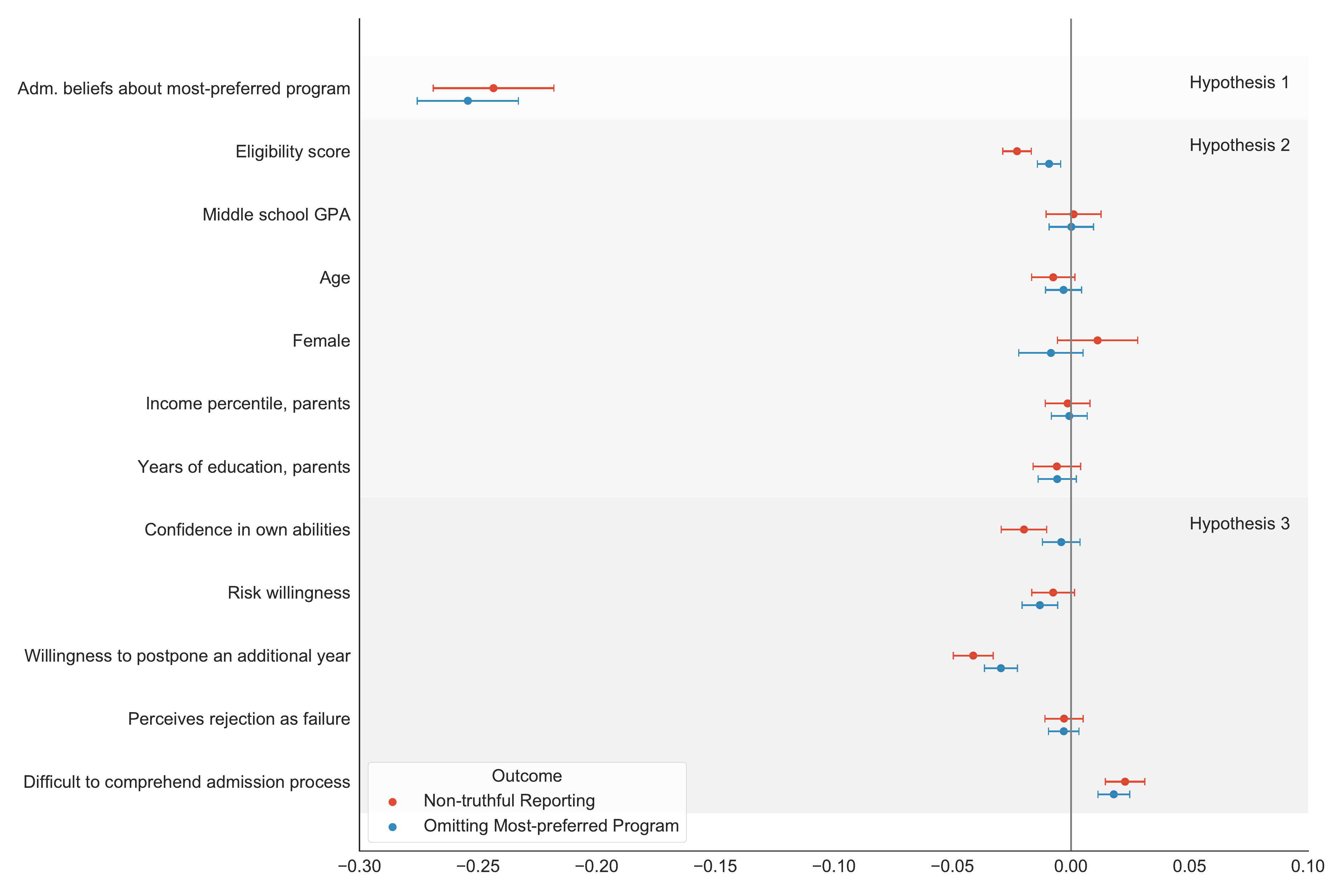}
    \end{subfigure}
\caption{Individual characteristics and truthful behavior; testing Hypotheses \ref{hypothesis:beliefs}, \ref{hypothesis:SES}, and \ref{hypothesis:Personality}}
\label{fig:hyp_1_to_3_ols_no_fe}
\floatfoot{Notes: The figure contains estimated coefficients from model \eqref{eq:model_hyp_1_2_3} with 95\% confidence intervals. The dependent variable is one of two measures of non-truthful behavior, see the legend. Each entry on the vertical axis corresponds to an explanatory variable and the horizontal axis measures the variable's coefficient size.
Standard errors of coefficients use robust estimation but are not clustered.}
\end{figure}

We proceed to estimate a model that includes other individual characteristics in the model of non-truthful behavior.
Figure \ref{fig:hyp_1_to_3_ols_no_fe} shows the estimated coefficients of non-truthful reporting and omitting most-preferred program as outcomes from model \eqref{eq:model_hyp_1_2_3}. Once more, we observe that the probability of non-truthful reporting and omitting most-preferred program decreases gradually with students' admission beliefs about their most-preferred program. This evidence supports Hypothesis~\ref{hypothesis:beliefs} even after controlling for additional explanations.

\paragraph{Individual characteristics and non-truthful behavior}
We proceed to examine Figure \ref{fig:hyp_1_to_3_ols_no_fe} for the coefficients for other individual characteristics. We find mixed support for Hypothesis~\ref{hypothesis:SES}. Only eligibility score has a significant estimate and in general, the coefficient sizes are relatively small compared to admission beliefs. This finding is a pretty stark contrast to the correlations observed in Figure~\ref{fig:correlation_non_truthful_explanatory}.

Finally, the figure suggests that students' personalities influence their decisions to report non-truthfully and omit their most-preferred program. That is, we do observe support for Hypothesis~\ref{hypothesis:Personality}. Students' confidence in own abilities and their willingness to postpone education an additional year decrease the probability of non-truthful reporting. Students who find it difficult to comprehend the admission process are more likely to report non-truthfully and omit their most-preferred program. 
However, by comparing the relative sizes of coefficients, personality seems to play a smaller role compared to admission beliefs. In conclusion, admission beliefs seem to be a leading explanation compared to the alternative explanations.

In our above results, we do not consider the consequences of truthful reporting. In the following section, we proceed to analyze who make a payoff-relevant omission. 

%% file: paper_sections/5_payoff_relevant_mistakes_analysis.tex
In this section, we investigate three potential explanations of payoff-relevant omissions. Specifically, we consider the following hypotheses \ref{hypothesis_mistake:beliefs_errors}, \ref{hypothesis_mistake:SES}, and \ref{hypothesis_mistake:Personality}. However, in order to analyze these, we require a measure for pessimistic admission beliefs. 

\paragraph{Measure of pessimistic admission beliefs} We define students' belief errors according to the same approach as \citet{Kapor2020HeterogeneousMechanisms, Agarwal2019RevealedModels}. We define student $i$'s belief error for admission at her most-preferred program $j$ to be the difference between her stated admission belief and the admission probability under rational expectations. Formally given by:
\begin{align}
    \textit{belief error}_{ij} = \hat{p}_{ij} - p_{ij}  
\end{align}
Thus, a student is pessimistic (optimistic) if $\textit{belief error}_{ij}$ is negative (positive). To calculate $p_{ij}$, admission probabilities under rational expectations, we apply an approach suggested by \citet{Agarwal2018DemandMechanism, Kapor2020HeterogeneousMechanisms, Larroucau2019DoProblem}. For each admission year, we draw a sample of applicants from the applicant population iid with replacements. For each sample draw, we solve for market-clearing cutoffs by running the assignment mechanism.\footnote{We apply an almost identical algorithm to the official algorithm used by the Ministry of Education. Our algorithm can reproduce the official matching between students and programs with 99.7\% accuracy.} We repeat this procedure 100 times, thus providing us with 100 simulated cutoff levels for each study program. By comparing a student's eligibility score to simulated cutoff levels, we obtain a student's admission probabilities under rational expectations. For an intuitive explanation, consider a student who has an eligibility score of 9. 70 out of 100 of the simulated cutoff levels of her most-preferred program are below 9. This implies, that her admission probabilities under rational expectations is 70\%.    

\begin{figure}[!ht]
    \begin{subfigure}[b]{.7\linewidth}
        \includegraphics[width=\linewidth]{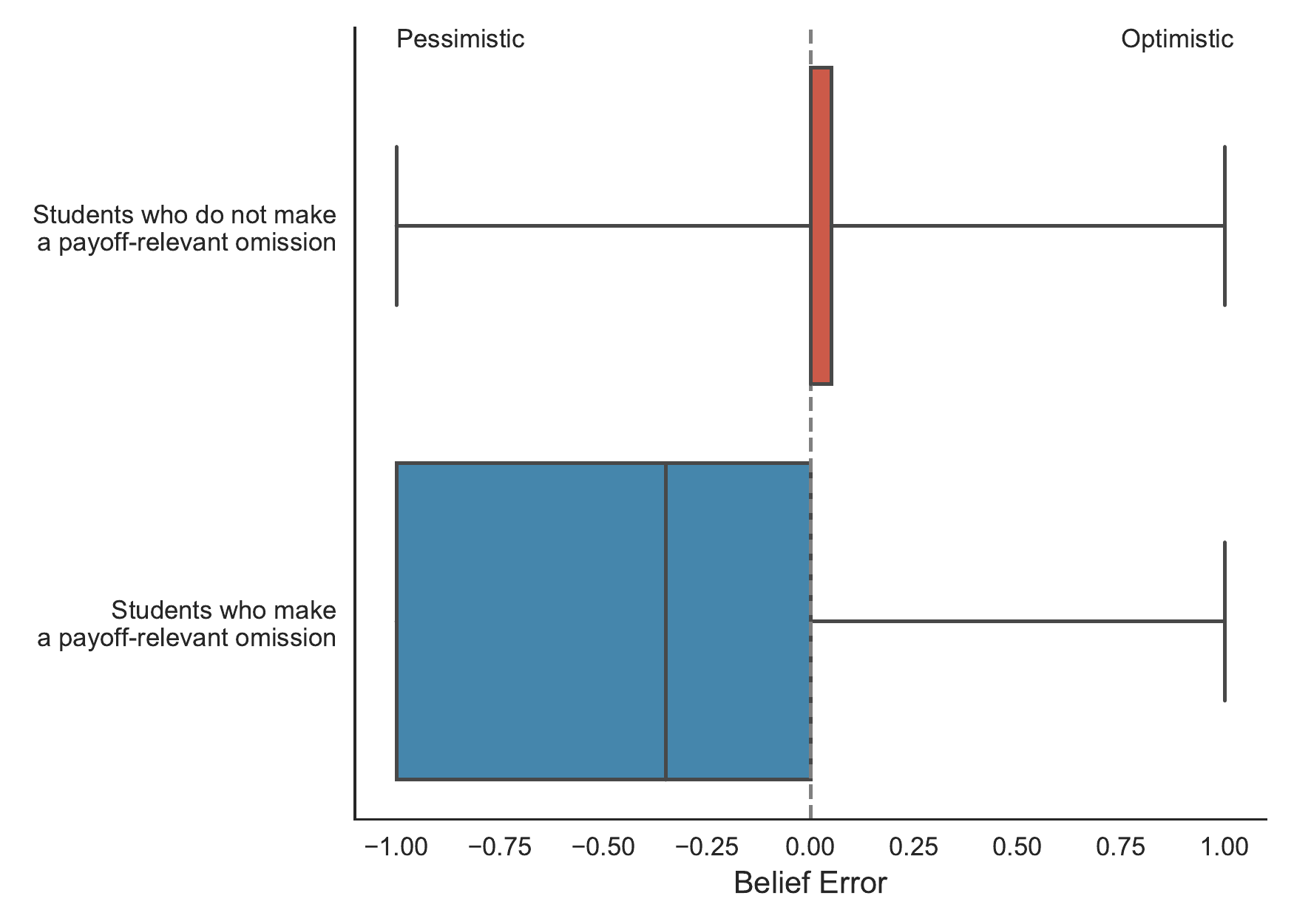}
    \end{subfigure}
\caption{Boxplot Distribution of Belief Errors}
\label{fig:belief_error_dist}
\end{figure}

We calculate the $\textit{belief error}_{ij}$ for the students who omit their most-preferred program. As descriptive evidence for Hypothesis~\ref{hypothesis_mistake:beliefs_errors}, Figure~\ref{fig:belief_error_dist} shows the distribution of beliefs errors conditional on whether the student made a payoff-relevant omission. From this figure, we observe that students who make a payoff-relevant omission are much more pessimistic than students who do not make a payoff-relevant omission. 

\begin{figure}[!ht]
    \begin{subfigure}[b]{.8\linewidth}
        \includegraphics[width=\linewidth]{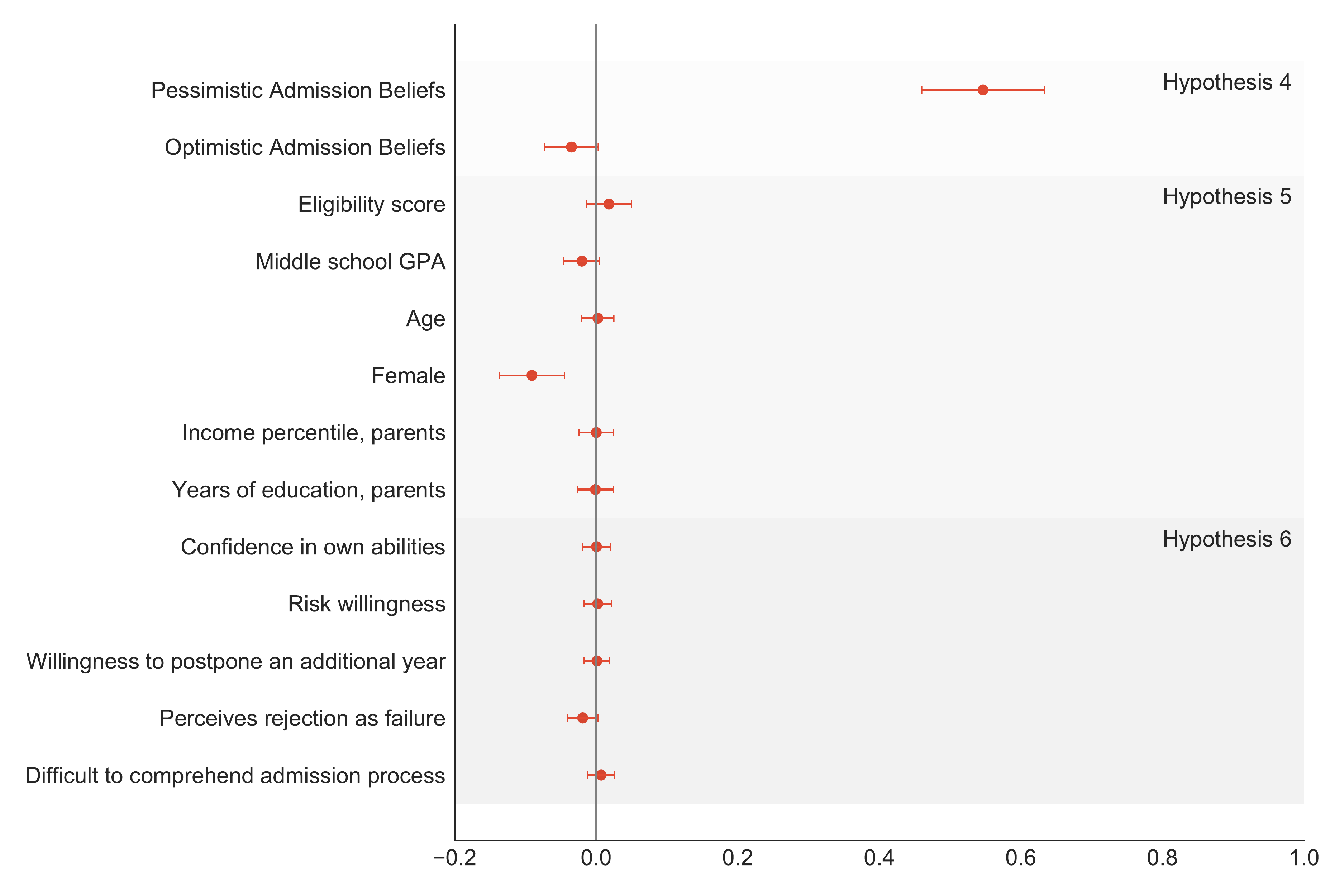}
    \end{subfigure}
\caption{Payoff-relevant Omission of Most-preferred Program}
\label{fig:regret_skip}
\floatfoot{Notes: The figure above shows coefficients from model (\ref{eq:model_hyp_4_5_6}). 
Coefficients are plotted with 95\% confidence intervals.
The dependent variable is payoff-relevant omission of most-preferred program. Each entry on the vertical axis corresponds to an explanatory variable and the horizontal axis measures the variable's coefficient size.
Standard errors of coefficients use robust estimation but are not clustered.}
\end{figure}

Figure \ref{fig:regret_skip} shows the estimated coefficients of model \ref{eq:model_hyp_4_5_6}. First, we observe that pessimistic admission beliefs drastically increase the risk of omitting the most-preferred program with payoff relevance, as is to be expected. All other coefficients are small and with substantially less statistical significance. Notably, female students are less likely to make a payoff-relevant omission of their most-preferred program. In the next section, we move on to measure how our presented evidence of non-truthful reporting affect the demand estimations under the assumption of truthful reporting.

%% file: paper_sections/6_demand_estimation_analysis.tex
In this section, we investigate how non-truthful reporting affects educational demand estimations. Specifically, we examine whether the assumption of truthful reporting can cause an estimation bias if a subgroup of students do not report truthfully. Moreover, we examine an alternative assumption to truthful reporting suggested by \citet{Fack2019BeyondAdmissions}. We leverage the fact that we observe survey respondents stated most-preferred program as well as their revealed most-preferred program. This allows us to estimate demand parameters using stated choices as well as revealed choices. Thus, by comparing the estimated demand parameters between these two models, we obtain a measure of estimation biases by using the truthful reporting assumption for educational demand estimation. In addition, we compare the estimated parameters from the alternative assumption to examine whether this alternative can reduce the estimation bias. 

We follow a standard random utility model approach and apply single choice conditional logit regressions for the estimation part.\footnote{We cannot apply an exploded conditional logit model because survey respondents only reveal their true top ranked choice.} The random utility model is specified as:

\begin{align}
    U_{i,j} = \theta_j - d_{i,j} + \gamma X_{i,j} + \epsilon_{i,j}
\end{align}

\noindent where $\theta_j$ are fixed effects of program $j$, and $d_{i,j}$ is the distance between student $i$'s postcode and the address of the university where program $j$ is located. We follow a standard procedure to localize the coefficient of $d_{i,j}$ to -1. $\Theta_j$ contains program fixed effects. $X_{i,j}$ contains three student-program-specific factors: student $i$'s high school GPA interacted with the average high school GPA among students in program $j$ denoted as \textit{Program Peer Quality}, the share of students with the same gender as student $i$ in program $j$ denoted as \textit{Same Gender Share}, and finally student $i$'s parents' income rank interacted with the average income rank among students' parents in program $j$ denoted as \textit{Program Peer Parents Inc}. The variables in $X_{i,j}$ are chosen based on their policy relevance and their varying correlation with non-truthful reporting. Specifically, we expect \textit{Program Peer Quality} to be highly correlated with non-truthful reporting because programs with high peer quality are selective and thus more likely to be omitted. 

Our benchmark model applies students' revealed most-preferred program as choice outcomes. That is, for students' who omit their most-preferred program, we use the choice they reveal in the survey to be their true top choice. We define this as \textit{Revealed Preferences}. Our first comparison model applies students' stated choice and thus assumes truthful reporting. Finally, our second comparison model also applies students' stated choices. However, it relaxes the truthful assumption by constraining students' choice set. Specifically, the choice set only considers feasible choices. A feasible choice is defined as a program with a cutoff level below the student's eligibility score. For example, if a student's eligibility score is 9, the regression model only considers programs with a cutoff below 9 in her choice set. This approach corresponds to a conditional logit model with a constrained choice set where one assumes the matching outcome is stable. We refer to \citet{Fack2019BeyondAdmissions} for a thorough explanation of this approach. 

Table \ref{tab:demand_est_offset} shows the estimated demand coefficients of the three models. Most notable is the difference in the coefficient sizes for \textit{Program Peer Quality}. When the truthful assumption is applied, the coefficient size is twice as large compared to the benchmark of revealed preferences. This suggests a potential bias in the coefficient estimate of \textit{Program Peer Quality}. Interestingly, this bias decreases if stability is used as an assumption instead of truthful reporting. 

\input{tables/clogit_demand_reg_offset}

%% file: tables/clogit_demand_reg_offset.tex
\begin{table}[htbp]
\centering
\small
\def\sym#1{\ifmmode^{#1}\else\(^{#1}\)\fi}
\begin{tabular}{l*{3}{c}}
\hline\hline
            &\multicolumn{1}{c}{(1)}&\multicolumn{1}{c}{(2)}&\multicolumn{1}{c}{(3)}\\
            &\multicolumn{1}{c}{Revealed Pref}&\multicolumn{1}{c}{Stated Pref}&\multicolumn{1}{c}{Stated Pref}\\
\hline
                        &            &            &            \\
Distance (1000 km)      &          -1&       -1&      -1\\
                        &           &      &      \\
Program Peer Quality    &        9.24&       19.32&        7.24\\
                        &      (0.32)&      (0.42)&      (0.44)\\
Same Gender Share       &        2.25&        2.24&        2.29\\
                        &      (0.06)&      (0.06)&      (0.06)\\
Program Peer Parents Inc&        5.66&        6.58&        7.50\\
                        &      (0.62)&      (0.63)&      (0.60)\\
\hline
N                       &   6,888,528&   6,888,528&   5,973,743\\
Revealed Preferences    &         Yes&          No&          No \\
Truthful Assumption     &          No&          Yes&         No \\
Stability Assumption    &          No&          No&         Yes\\
\hline\hline
\end{tabular}
\caption{Comparison of demand models}
\label{tab:demand_est_offset}
\floatfoot{Notes: This table provides regression coefficients from estimating demand models under the three approaches of (1) revealed preferences, (2) baseline model, and (3) using the stability assumption, see text for details. The models are estimated with standard errors clustered at the study program level.}
\end{table}

%% file: paper_sections/robustness_analyses.tex
In this section, we conduct robustness analyses of our main results related to admission beliefs and non-truthful reporting. We perform the following robustness analyses: i) inclusion of admission beliefs about an alternative admission channel, ii) an interplay between Hypothesis \ref{hypothesis:beliefs} with Hypotheses \ref{hypothesis:SES}, and \ref{hypothesis:Personality}, iii) examination of whether our results are driven by students failing to understand strategy-proofness, iv) and finally different specifications of the model used to test Hypotheses \ref{hypothesis:beliefs}-\ref{hypothesis:Personality} and \ref{hypothesis_mistake:beliefs_errors}-\ref{hypothesis_mistake:Personality}.   

\paragraph{Alternative admission quota}
The Danish admission system allows study programs to admit a minor share of students through alternative evaluation criteria than eligibility scores. In our main analysis, we omit this alternative path to admission. However, we conduct a robustness analysis to show that omitting the alternative path does not affect our results for Hypotheses \ref{hypothesis:beliefs}-\ref{hypothesis:Personality}. Our robustness analysis expands the measure of stated admission beliefs about the most-preferred program by including the stated admission beliefs about the alternative path of admission. We refer to the latter as the combined admission beliefs about the most-preferred program, which is formally given by:

\begin{align}
    \hat{p}_{\textit{combined}} = \hat{p}_{\textit{eligibility score}}+(1-\hat{p}_{\textit{eligibility score}})*\hat{p}_{\textit{alternative criteria}}
\end{align}

\noindent where $\hat{p}_{\textit{eligibility score}}$ is the stated admission beliefs about admission to a student's most-preferred program by eligibility score, and $\hat{p}_{\textit{alternative criteria}}$ is the stated admission beliefs about admission to a student's most-preferred program by alternative criteria. We estimate two separate models using the two measures of admission beliefs and compare the coefficient results. Table \ref{tab:robust_reg_2021} contains the estimated coefficients. The coefficient of combined admission beliefs about the most-preferred program remains negative and it is not significantly different to the coefficient of admission beliefs about the most-preferred program.

\paragraph{Interplay of admission belief and student characteristics}
Theory from economic models predicts that students adapt their reported preferences and truthfulness according to beliefs. Yet, the models assume that this adaptation in reporting is identical across individuals. E.g., a student's adaption is unaffected by her demographics. To explore this theory, we therefore investigate whether individual or contextual factors matter for students' adaptations. 
We operationalize this by exploring the possibility of an interplay between Hypothesis \ref{hypothesis:beliefs} with Hypothesis \ref{hypothesis:SES} and Hypothesis \ref{hypothesis:Personality}. We consider the following hypothesis. 

\begin{hypothesis}
\label{hypothesis:HE_beliefs} Students' demographics, SES, academic achievements, and personality affect students' propensity to report non-truthfully and omit most-preferred programs if they believe the probability of admission into their most-preferred program is zero or very low.
\end{hypothesis}

To test Hypothesis~\ref{hypothesis:HE_beliefs}, we expand our main model with an interaction between our measures of students' characteristics (e.g., gender) and their stated admission beliefs:
\begin{align}
    y_i = \alpha + \beta X_{H1}+\Gamma X_{H2}+\Delta X_{H3}+ \zeta (X_{H1} \times X_{H2,H3}) +\Theta+\epsilon_i
    \label{eq:model_hyp_1_2_3_7}
\end{align}

\begin{figure}[!ht]
    \begin{subfigure}[b]{.9\linewidth}
        \includegraphics[width=\linewidth]{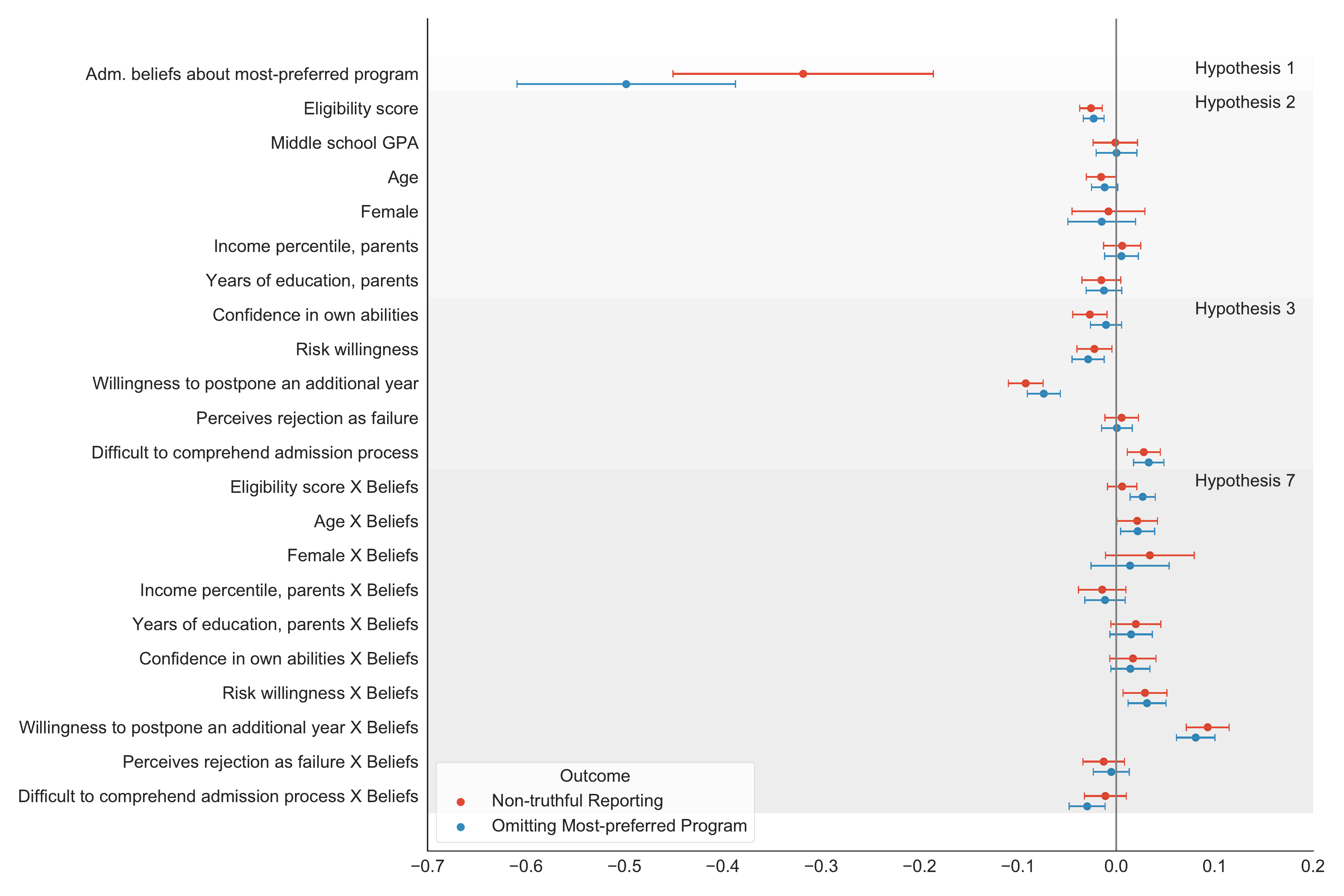}
    \end{subfigure}
\caption{Coefficient Estimates to Test Hypothesis \ref{hypothesis:HE_beliefs}}
\label{fig:robust_HE}
\floatfoot{Notes: The figure contains estimated coefficients of model (\ref{eq:model_hyp_1_2_3_7}). 
Coefficients are plotted with 95\% confidence intervals.
 The dependent variable is one of two measures of non-truthful behavior, see the legend. Each entry on the vertical axis corresponds to an explanatory variable and the horizontal axis measures the variable's coefficient size.
Standard errors of coefficients use robust estimation but are not clustered.}
\end{figure}

\noindent We present the estimated coefficients in Figure~\ref{fig:robust_HE}. The coefficients show that the association between admission beliefs and non-truthful reporting is moderated by a number of the student characteristics. Students' willingness for risk and postponing admission as well as their eligibility score lower this association. Conversely, difficulty in comprehending the admission process increases the association. It appears that the factors, students' age, length of parental education and confidence in own abilities also lower the association, but we are unable to distinguish these from null coefficients. 

An example of what the estimates entail is that among students who believe their admission probability is zero, or close to zero, the propensity to report non-truthfully or omit the most-preferred program is lower for risk seeking students compared to risk averse students. However, among students who believe their admission probability is very certain, the propensity to report non-truthfully or omit the most-preferred program is the same among risk seeking and risk averse students.

\paragraph{Understanding of strategy-proofness}
We proceed by exploring an alternative explanation to Hypotheses \ref{hypothesis:beliefs}-\ref{hypothesis:Personality}, which is that students do not fully understand the mechanism:

\begin{hypothesis}
\label{hypothesis:fail_strategyproof} Students report non-truthfully and omit most-preferred programs because they fail to understand that the admission mechanism is strategy-proof.
\end{hypothesis}

To test Hypothesis \ref{hypothesis:fail_strategyproof}, we use answers to the following question presented to survey respondents in the 2021 wave: \textit{Consider the following statement about the importance of a program's ranking for admission. Imagine that a student, Adam, has applied to a public school teacher program as his most-preferred program and social worker program as his second-most-preferred program. Now imagine that Adam is not offered a place in his most-preferred program. Would Adam's probability of being to the social worker program change in the counterfactual event where he omits the teacher program and instead applies to social worker as his most-preferred program?} The answer possibilities are: (1) \textit{No, it would remain the same}, (2) \textit{Yes, the admission probability would increase}, (3) \textit{Yes, the admission probability would decrease}, and (4) \textit{I don't know}. The correct answer is (1) since the admission is strategy-proof. Answer (2) to (4) are incorrect. We create an indicator variable \textit{Understands strategy-proofness} which is 1 if a student answers \textit{No, it would remain the same}. The indicator variable is then included in our regression model. Under Hypothesis  \ref{hypothesis:fail_strategyproof} we would expect the coefficient estimate to be negative. In Table~\ref{tab:robust_reg_2021}, we do indeed observe a negative estimate. However, the estimates are statistically insignificant for non-truthful reporting and borderline significant for omitting most-preferred program. Moreover, the inclusion of Hypothesis \ref{hypothesis:fail_strategyproof} does not affect our conclusion about Hypothesis \ref{hypothesis:beliefs}. Thus, we do not consider failing to understand strategy-proofness to be a major explanation for non-truthful reporting in our context. 

\input{tables/linreg_robustness_2021_survey}

\paragraph{Observable selection - adjusting for selected participation} We verify that our results are robust to estimation where we upweigh data from respondents who tend to participate less. We do this by re-estimating our models with Weighted Least Squares. \cite{Dutz2021SelectionSurveys} find that weighting approach may give slightly more precise estimates when no incentives are given for participating in surveys. We include the results from estimating models \ref{eq:model_hyp_1_2_3} and \ref{eq:model_hyp_4_5_6} using WLS instead of OLS in Appendix \ref{app:results_WLS}. None of our results are affect by this approach.

\paragraph{Unobservable selection - estimation with program fixed effects}
Another worry is that students who participated in the survey differ in unobserved ways from the sampled population and that this affects our estimates. To gauge this, we estimate our models of truthful behavior where we include fixed effects at the level of most preferred study programs. This captures unobserved selection into fields and programs.
We have included the results from estimating the models in Appendix~\ref{app:results_FE}. The inclusion of fixed effects does not change our conclusions of Hypotheses 1-6, see, e.g., Figures \ref{fig:hyp_1_to_3_ols_fe}, and
\ref{fig:model_hyp_4_5_6_fe} that are copies of figures in the main text estimated with study program level fixed effects.


\paragraph{Investigating determinants of non-truthfulness separately}
Our final robustness analysis tests Hypotheses \ref{hypothesis:beliefs}-\ref{hypothesis:Personality} separately. This allows us to better understand the interplay between the hypotheses, and compare our results to studies on non-truthful reporting which do not have as rich data. Specifically, we estimate the following three models:
\begin{align}
    y_i = \alpha + \beta X_{H1}+\Theta+\epsilon_i
\end{align}
\begin{align}
    y_i = \alpha + \Gamma X_{H2}+\Theta+\epsilon_i
\end{align}
\begin{align}
    y_i = \alpha + \Delta X_{H3}+\Theta+\epsilon_i
\end{align}
where outcomes and explanatory variables are identical to those mentioned in our main regression model.\footnote{The only exception is $X_{H1}$ which is estimated as a continuous variable in the robustness analysis} The results are shown in Tables \ref{tab:lin_reg_res_FE_nontruthful_2021} and \ref{tab:lin_reg_res_FE_skipimp_2021}. We observe that female students and students with low confidence in their own abilities are associated with a significantly higher rate of non-truthful reporting and omitting of most-preferred program when we do not confound on alternative explanations.

\input{tables/linreg_FE_nontruthful_2021}
\input{tables/linreg_FE_skipimp_2021}

%% file: tables/linreg_robustness_2021_survey.tex
\begin{table}[]
\centering
\small
\renewcommand{\arraystretch}{1.25} 
{
\def\sym#1{\ifmmode^{#1}\else\(^{#1}\)\fi}
\begin{tabular}{l*{4}{p{2cm}}}
\hline\hline
 & \multicolumn{2}{l}{\textbf{Reports}}  & \multicolumn{2}{l}{\textbf{Omits}} \\
  & \multicolumn{2}{l}{\textbf{non-truthfully}}  & \multicolumn{2}{l}{\textbf{most-preferred program}} \\
 & (1) & (2) & (3) & (4) \\  
\hline
\multicolumn{5}{l}{\textit{Hypothesis 1} } \\

Adm beliefs most-preferred program&      -0.253&            &      -0.237&            \\
            &     (0.025)&            &     (0.021)&            \\
Combined adm beliefs most-preferred program&            &      -0.310&            &      -0.295\\
            &            &     (0.028)&            &     (0.025)\\
\multicolumn{5}{l}{\textit{Hypothesis 2}} \\
Eligibility score&      -0.026&      -0.029&      -0.020&      -0.022\\
            &     (0.006)&     (0.005)&     (0.005)&     (0.004)\\
Middle school GPA&      -0.013&      -0.012&      -0.010&      -0.009\\
            &     (0.009)&     (0.009)&     (0.008)&     (0.008)\\
Age         &      -0.015&      -0.014&      -0.006&      -0.005\\
            &     (0.007)&     (0.007)&     (0.006)&     (0.006)\\
Female      &       0.036&       0.033&       0.023&       0.020\\
            &     (0.015)&     (0.015)&     (0.012)&     (0.012)\\
Income percentile, parents&      -0.006&      -0.007&      -0.014&      -0.014\\
            &     (0.007)&     (0.007)&     (0.006)&     (0.006)\\
Years of education, parents&      -0.002&      -0.001&      -0.001&      -0.000\\
            &     (0.008)&     (0.007)&     (0.006)&     (0.006)\\
\multicolumn{5}{l}{\textit{Hypothesis 3}} \\
Confidence in own abilities&      -0.003&      -0.001&       0.007&       0.009\\
            &     (0.007)&     (0.007)&     (0.006)&     (0.006)\\
Risk willingness&      -0.020&      -0.020&      -0.025&      -0.025\\
            &     (0.007)&     (0.007)&     (0.006)&     (0.006)\\
Willingness to postpone an additional year&      -0.032&      -0.030&      -0.023&      -0.021\\
            &     (0.007)&     (0.007)&     (0.006)&     (0.005)\\
Perceives rejection as failure&      -0.009&      -0.011&      -0.008&      -0.009\\
            &     (0.006)&     (0.006)&     (0.005)&     (0.005)\\
Difficult to comprehend admission process&       0.023&       0.021&       0.017&       0.015\\
            &     (0.006)&     (0.006)&     (0.005)&     (0.005)\\
\multicolumn{5}{l}{\textit{Hypothesis 8} } \\
Understands strategy-proofness.&       -0.025&       -0.025&    -0.028   &       -0.028\\
            &     (0.015)&     (0.014)&     (0.012)&     (0.012)\\

\hline
N           &    3,787        &    3,787      &    3,787       &    3,787         \\
Study program fixed effects&         No         &         No         &         No         &         No         \\
\hline\hline
\end{tabular}
}
\caption{Coefficient Estimates of Robustness Analyses}
\label{tab:robust_reg_2021}
\floatfoot{Notes: Outcome variable of equations (1) and (2) is non-truthful reporting. Outcome variable of equations (3) and (4) is omitting most-preferred program. Robust standard errors are reported in parenthesis. The models are estimated on survey respondents from the 2021 wave, since the 2020 wave did not include the required questions.}

\end{table}

%% file: tables/linreg_FE_nontruthful_2021.tex
\begin{table}[]
\centering
\small
\renewcommand{\arraystretch}{1.25} 
{
\def\sym#1{\ifmmode^{#1}\else\(^{#1}\)\fi}
\begin{tabular}{l*{4}{c}}
\hline\hline
 & \multicolumn{4}{c}{\textbf{Reports non-truthfully}} \\
 & (1) & (2) & (3) & (4) \\                               
\hline
\multicolumn{5}{l}{\textit{Hypothesis 1} } \\
Adm beliefs most-preferred program&      -0.299&            &            &      -0.243\\
            &     (0.011)&            &            &     (0.013)         \\
\textit{Hypothesis 2}                             &                 &                 &               &          \\            
Eligibility score&            &      -0.048&            &      -0.023\\
            &            &     (0.003)&            &     (0.003)\\
Middle school GPA&            &       0.001&            &      -0.001\\
            &            &     (0.006)&            &     (0.006)\\
Age         &            &      -0.006&            &      -0.008\\
            &            &     (0.005)&            &     (0.005)\\
Female      &            &       0.050&            &       0.011\\
            &            &     (0.009)&            &     (0.009)\\
Income percentile, parents&            &      -0.002&            &      -0.001\\
            &            &     (0.005)&            &     (0.005)\\
Years of education, parents&            &      -0.009&            &      -0.006\\
            &            &     (0.005)&            &     (0.005)\\
\textit{Hypothesis 3}                              &                 &                 &               &          \\                 
Confidence in own abilities&            &            &      -0.055&      -0.020\\
            &            &            &     (0.005)&     (0.005)\\
Risk willingness&            &            &       0.025&      -0.008\\
            &            &            &     (0.005)&     (0.005)\\
Willingness to postpone an additional year&            &            &      -0.040&      -0.041\\
            &            &            &     (0.005)&     (0.004)\\
Perceives rejection as failure&            &            &       0.010&      -0.003\\
            &            &            &     (0.004)&     (0.004)\\
Difficult to comprehend admission process&            &            &       0.025&       0.023\\
            &            &            &     (0.004)&     (0.004)\\
\hline
N           &    8,049         &    8,049        &    8,049         &    8,049        \\
Study program fixed effects&         No         &         No         &         No         &         No         \\
Survey wave fixed effects &         Yes         &         Yes         &         Yes         &         Yes         \\
\hline\hline
\end{tabular}
}
\caption{Coefficient Results of Models for Non-truthful Reporting}
\label{tab:lin_reg_res_FE_nontruthful_2021}
\floatfoot{Notes: The table shows results for four separate linear probability regression models. The table reports coefficient estimates and robust standard errors in parenthesis. Each equation considers non-truthful reporting as an outcome variable.}

\end{table}

%% file: tables/linreg_FE_skipimp_2021.tex
\begin{table}[]
\centering
\small
\renewcommand{\arraystretch}{1.25} 
{
\def\sym#1{\ifmmode^{#1}\else\(^{#1}\)\fi}
\begin{tabular}{l*{4}{c}}
\hline\hline
 & \multicolumn{4}{c}{\textbf{Omits Most-preferred Program}} \\
 & (1) & (2) & (3) & (4) \\  
\hline
\multicolumn{5}{l}{\textit{Hypothesis 1} } \\
Adm beliefs most-preferred program&      -0.271&            &            &      -0.254\\
            &     (0.009)&            &            &     (0.011)        \\
\textit{\textit{Hypothesis 2}}                             &                 &                 &               &          \\
Eligibility score&            &      -0.033&            &      -0.009\\
            &            &     (0.002)&            &     (0.003)\\
Middle school GPA&            &      0.001&            &      -0.000\\
            &            &     (0.005)&            &     (0.005)\\
Age         &            &       0.005&            &      -0.003\\
            &            &     (0.004)&            &     (0.005)\\
Female      &            &       0.026&            &       0.008\\
            &            &     (0.007)&            &     (0.007)\\
Income percentile, parents&            &      0.000&            &      -0.001\\
            &            &     (0.004)&            &     (0.004)\\
Years of education, parents&            &      -0.008&            &      0.004\\
            &            &     (0.004)&            &     (0.004)\\
\textit{Hypothesis 3}                             &                 &                 &               &          \\  
Confidence in own abilities&            &            &      -0.032&      -0.004\\
            &            &            &     (0.004)&     (0.004)\\
Risk willingness&            &            &       0.015&      -0.013\\
            &            &            &     (0.004)&     (0.004)\\
Willingness to postpone an additional year&            &            &      -0.028&      -0.030\\
            &            &            &     (0.004)&     (0.004)\\
Perceives rejection as failure&            &            &       0.007&      -0.003\\
            &            &            &     (0.004)&     (0.003)\\
Difficult to comprehend admission process&            &            &       0.020&       0.018\\
            &            &            &     (0.004)&     (0.003)\\
\hline
N           &    8,049         &    8,049        &    8,049         &    8,049        \\
Study program fixed effects&         No         &         No         &         No         &         No         \\
Survey wave fixed effects &         Yes         &         Yes         &         Yes         &         Yes         \\
\hline\hline
\end{tabular}
}
\caption{Coefficient Results of Models for Omitting Most-preferred Program}
\label{tab:lin_reg_res_FE_skipimp_2021}
\floatfoot{Notes: The table shows results for four separate linear probability regression models. The table reports coefficient estimates and robust standard errors in parenthesis. Each equation considers omitting most-preferred program an outcome variable.}

\end{table}

%% file: paper_sections/literature_review.tex
In this section, we review literature related to findings and causes of non-truthful reporting and econometric methods for demand estimation using truthful reporting and alternative assumptions. 

Research within non-truthful reporting faces a fundamental challenge in  identifying non-truthful reporting because researchers do not observe students' true preferences. Typical approaches to overcome this challenge are lab experiments, surveys, or exploiting a unique situation with objectively dominated reports. An example of the latter is \citet{Hassidim2016StrategicEnvironment} who consider students applying to psychology programs under a strategy-proof admission system. The students can list a preferred study program with and without a scholarship. The authors argue that a scholarship position is unambiguously preferred to a non-scholarship position. Hence, if students omit a scholarship position, they play a dominated strategy. They observe that $19.5\%$ of the students misreport by omitting a scholarship option in their reported preferences. Using a similar approach, \citet{Shorrer2022,Shorrer2023} study a Deferred Acceptance mechanism used for centralized admission of higher education in Hungary. They find that about $10\%$ of students omit a free of cost option of a tuition waiver. What is more, up to $12\%$ of the non-truthful reporting is costly because some students were in fact eligible for a tuition waiver. \citet{Artemov2021StrategicResearch} examine a serial dictatorship used for Tertiary Admission in Victoria, Australia. They show that $34\%$ of students who applied for a position with a tuition waiver available, omit the tuition waiver option in their reported preferences. However, they show that the majority of non-truthful reporting is payoff irrelevant.  
\citeauthor{Chen2019Self-selectionChoice} (\citeyear{Chen2019Self-selectionChoice}) use a survey approach to elicit the true preferences of a sub-group of students applying to high schools in Mexico City. When they compare the actual reported preferences to the surveyed preferences, they find that about $20\%$ of the students omit reporting very selective high schools.\footnote{The authors define this behavior as self-selection} Moreover, they find that approximately $23\%$ of the students who intentionally omit selective high schools would have been admitted to a selective high school had they not omitted it. \citeauthor{Rees-Jones2018SuboptimalMatch} (\citeyear{Rees-Jones2018SuboptimalMatch}) also uses a survey to elicit preferences of graduating medical students who apply for a position through the National Resident Matching Program. His analysis shows that $17\%$ of the students self-assess their reported preferences to be non-truthful. Finally, \citet{Larroucau2019DoProblem} survey Chilean applicants for higher education and report that 57\% report non-truthfully.\footnote{They separate non-truthful reporting between applicants who submit a complete ROL and those who do not. The non-truthful rate of 57\% is among students who do not submit a complete ROL.} As for the evidence of non-truthful reporting in lab experiments, \citet{Hakimov2021ExperimentsSurvey} provide a review of non-truthful reporting in lab experiments. The share of truth-telling reports is in the range of $50\%$ to $80\%$ for strategy-proof mechanisms.\footnote{The strategy-proof mechanisms considered are Deferred Acceptance and Top Trading Cycles}. \citet{Rees-Jones2018AnMatch} conduct an experiment to investigate non-truthful reporting in residency match of medical students. Similar to our findings, they also observe a higher non-truthful rate among students with lower priority scores. \\

To explain these findings, researchers propose a variety of explanations. It is suggested that non-truthful reporting is an unintentional behavior caused by students who fail to identify honesty as a dominant strategy \citep{Hassidim2017TheYou}. This explanation is supported by higher rates of non-truthful reporting among students with low cognitive abilities. However, evidence from lab experiments show that students continue to misreport even after being clearly instructed about the existence of an optimal strategy in strategy-proof mechanisms \citep{Hakimov2021ExperimentsSurvey}. This suggests that non-truthful reporting is in fact an intentional behavior. \citet{Rees-Jones2023BehavioralReview} explain how theories from behavioral economics can explain intentional non-truthful reporting. Alternatively, \citeauthor{Hassidim2017TheYou} (\citeyear{Hassidim2017TheYou}) address three potential explanations for intentional non-truthful reporting: mistrust, non-classical utility, and rational omitting. If students mistrust the advice given by a market maker, they may choose not to follow the recommended strategy. The explanation of mistrust is supported by lab evidence showing that rates of truthful reports decrease when students doubt the legitimacy of advice given to them prior to reporting preferences in a top trading cycle \citep{Guillen2014LyingMechanism}. The non-classical utility explanation covers multiple options: ego utility \citep{Koszegi2006EgoChoice}, altruistic motives \citep{Hassidim2017TheYou}, and expectation-based loss aversion \citep{Meisner2023LossMechanisms,Dreyfuss2022Expectations-BasedMechanisms}. Finally, rational omission of most-preferred programs is an explanation that relies on the proof that omitting schools with a perceived probability of zero can be a weakly dominating strategy in strategy-proof mechanisms \citep{Fack2019BeyondAdmissions,Chen2019Self-selectionChoice,Larroucau2019DoProblem} and a dominating strategy if students face a constrained ROL or application costs \citep{Haeringer2009ConstrainedChoice,Arslan2021PreferenceLists,Larroucau2019DoProblem,He2022ApplicationMarkets}. This explanation is supported by empirical evidence suggesting that non-truthful reporting is more likely to occur if non-truthful reporting is payoff-irrelevant \citep{Artemov2021StrategicResearch}. \\ 

We now proceed to review literature on econometric methods for demand estimation using truthful reporting and alternative assumptions. For the past two decades a growing body of research has applied application data from educational matching markets to estimate students' and families' preferences for education (e.g. \citet{Calsamiglia2020StructuralAlternatives,Abdulkadiroglu2017ResearchEvaluation,Burgess2015WhatChoice}). In situations where the matching markets are strategy-proof, researchers often assume students report truthfully (e.g., \citet{Abdulkadiroglu2021SchoolQuality}). However, as pointed out by \citet{Fack2019BeyondAdmissions}, strategy-proofness is a weakly dominating strategy, leaving open an issue of alternative equilibrium behavior which does not coincide with truthful reporting. They clarify the implication of truthful reporting and propose to use stability as an alternative identification assumption. They argue that stability is an attractive alternative because it does not require information about students' admission beliefs and can be estimated using standard choice models. Although stability is a less restrictive assumption, it nevertheless requires that students can perfectly predict post-matching cutoffs. \citet{Artemov2021StrategicResearch,Arslan2021PreferenceLists} both relax the stability assumption by proposing weaker-stable assumptions and develop methods that retain the computational attractiveness of stability. Finally, \citet{Larroucau2019DoProblem} try to solve the issue by considering students reporting as a portfolio problem in order to develop a sophisticated demand estimation method.        


%% file: paper_sections/discussion_and_conclusion.tex
In this paper, we demonstrate that a student's admission beliefs are a leading explanation for her decision to report non-truthfully in a strategy-proof admission system. Furthermore, we show that welfare losses from non-truthful reporting are primarily driven by students who are overly pessimistic about their admission prospects. Finally, we show that assuming truthful reporting for demand estimations generates biased estimates.  

Our findings contribute to a discussion of whether the scope for analyzing admission markets should be broader. Traditionally, mechanism design analyzes agents' behavior in a scope where preferences and information are well-established. Unquestionably, this scope has enabled remarkable research results that solve fundamental matching market problems. Nevertheless, this scope cannot easily explain why agents' observed behavior deviates from optimal strategies. E.g., the theoretical results for strategy-proofness suggest admission beliefs should not influence admission behavior. Our results therefore play an important role in explaining why standard theoretical results may not hold in real-life matchings. A growing body of literature acknowledges this challenge and therefore tries evaluate welfare implications of a broader scope that also considers agents' preference formation and information acquisition \citep{Arteaga2022SmartChoice,He2022ApplicationMarkets,Grenet2021DecentralizingAdmissions}. It is our conjecture that the reason why we observe limited welfare implications of non-truthful behavior is the high degree of transparency and stability in the Danish admission system. Specifically, students observe historic program cutoff levels and can, therefore, relatively easily assess whether a program is accessible to them. Consequently, we argue that policy makers can reduce welfare losses from non-truthful behavior by increasing the transparency of admissions. E.g., by implementing a smart tool as suggested by \citet{Arteaga2022SmartChoice}. 

Finally, our findings contribute to a discussion of whether to apply truthful reporting as an identification assumption in demand estimations. Our results suggest that the truthful assumption should be used with caution if the variable of interest is expected to be correlated with non-truthful behavior. To illustrate this point, we show how assuming truthful reporting causes a biased demand estimate of a programs' peer quality, because programs with high peer quality are frequently omitted by non-truthful students. As a consequence of our findings, we suggest that researchers should at least conduct robustness analyses with alternative assumptions if they choose to assume truthful reporting.

We note that we do not know the underlying structural equation model that governs non-truthful behavior. Our estimated coefficients of the determinants of non-truthful behavior and its consequences are estimated in a regression setup that eliminates all joint variation. However, this included the background factors of the parents' education and income that correlated positively with truthful behavior. As parents' background precedes the application decision, and all other outcomes we measure, one could argue that higher socioeconomic background leads to more truthful behavior. 

As a final note, we point out that our results may be influenced by survey response bias and selection bias. Our survey design is intended to mitigate typical response biases such as question order bias, demand characteristics, and social desirability bias. If our survey contains response bias, we consider it most likely that the bias is isolated to questions related to students' personality, e.g., students' self-stated confidence in their own abilities. However, we do not believe survey response bias would change the main conclusions of our paper. In terms of selection bias, we do identify observable selection on survey participation. Although the observable differences between our population and sample are statistically significant our robustness analysis show it does not affect our findings and conclusions. In conclusion, we consider our results to have external validity.

%% file: paper_sections/X_appendix.tex
\subsection{Summary of Danish admission to higher education}
\input{tables/admission_summary_stats}

\subsection{Detailed description of survey covariates} \label{app:survey_var_describe}

\noindent\textbf{Comprehension of admission process}: Agreement with the following statement: `It is generally difficult to comprehend how the admission process for higher education works. We create an indicator variable for the students who agree or completely agree with this statement.'

\noindent\textbf{Willingness to postpone}: Responds `Agree' or `Completely agree' to the statement:  `I am willing to wait another year, if I do not get admitted to any of the study programs I applied to this year'.

\noindent\textbf{Perceives rejection as a failure}: Responds `Agree' or `Completely agree' to the following statement: `It is a failure to apply for a study program and subsequently be rejected'.

\noindent\textbf{Confidence in own abilities}: Reported 0 to 10 rating to the following question: `Are you generally a person, who has large confidence in your own abilities?'. An answer of 0 corresponds to `I have very little confidence to my abilities'. An answer of 10 corresponds to `I have very high confidence to my abilities'. Thus, a higher number indicates higher self-stated confidence in own abilities.

\noindent\textbf{Risk profile}: Respondents between 0 to 10 rating to the following question: `Are you generally a person, who is willing to take risk or do you actively avoid risks?'. An answer of 0 corresponds to `I am not at all willing to take any risk'. An answer of 10 corresponds to `I am very willing to take risk'. Thus, a higher number indicates higher willingness to take risk.

\newpage

\newpage
\newpage    

\subsection{Model output from Weighted Least Squares estimation}\label{app:results_WLS}

\begin{figure}[!ht]
    \begin{subfigure}[b]{.9\linewidth}
        \includegraphics[width=\linewidth]{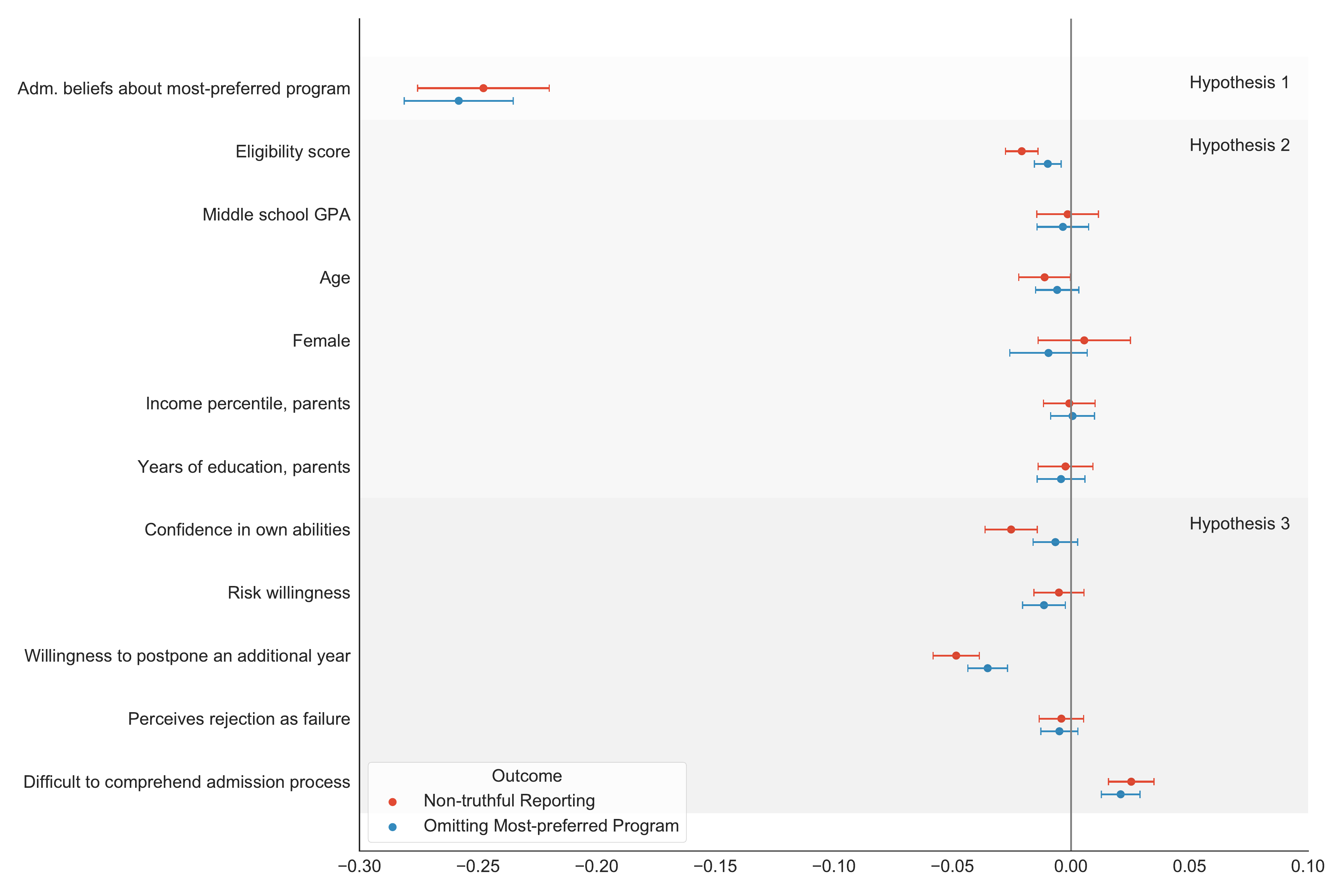}
    \end{subfigure}
\caption{Individual characteristics and truthful behavior; testing Hypotheses \ref{hypothesis:beliefs}, \ref{hypothesis:SES}, and \ref{hypothesis:Personality} using WLS}
\label{fig:hyp_1_to_3_wls_no_fe}
\floatfoot{Notes: The figure contains estimated coefficients from model \eqref{eq:model_hyp_1_2_3} with 95\% confidence intervals using WLS instead of OLS. The dependent variable is one of two measures of non-truthful behavior, see the legend. Each entry on the vertical axis corresponds to an explanatory variable and the horizontal axis measures the variable's coefficient size. Standard errors of coefficients use robust estimation but are not clustered.}
\end{figure}

\begin{figure}[!ht]
    \begin{subfigure}[b]{.8\linewidth}
        \includegraphics[width=\linewidth]{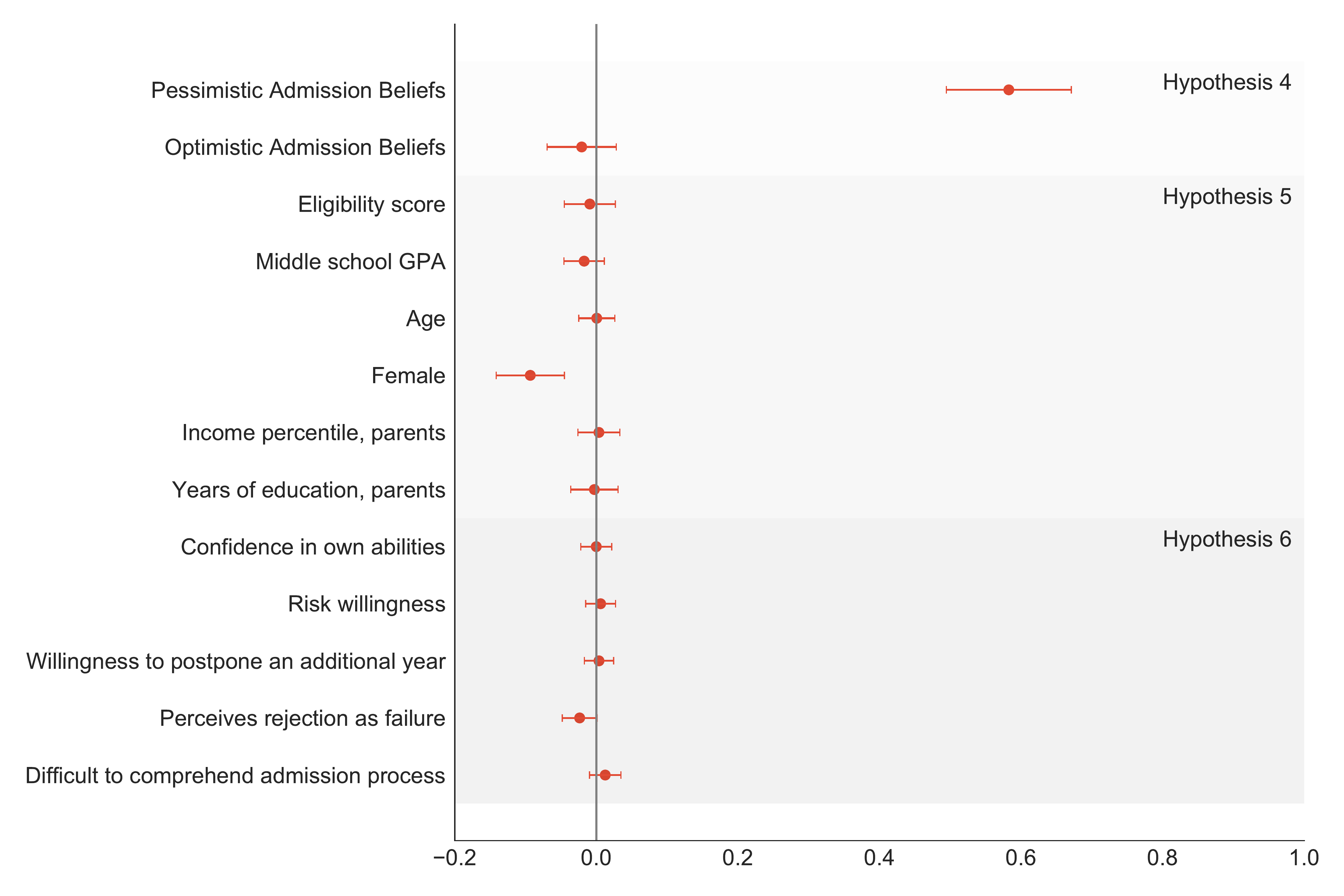}
    \end{subfigure}
\caption{Payoff-relevant Omission of Most-preferred Program using WLS}
\label{fig:hyp_4_to_6_wls_no_fe}
\floatfoot{Notes: The figure above shows coefficients from model (\ref{eq:model_hyp_4_5_6}) using WLS instead of OLS. Coefficients are plotted with 95\% confidence intervals. The dependent variable is payoff-relevant omission of most-preferred program. Each entry on the vertical axis corresponds to an explanatory variable and the horizontal axis measures the variable's coefficient size. Standard errors of coefficients use robust estimation but are not clustered.}
\end{figure}

\subsection{Model output with fixed effects at study program level}\label{app:results_FE}

\begin{figure}[!ht]
    \begin{subfigure}[b]{.9\linewidth}
        \includegraphics[width=\linewidth]{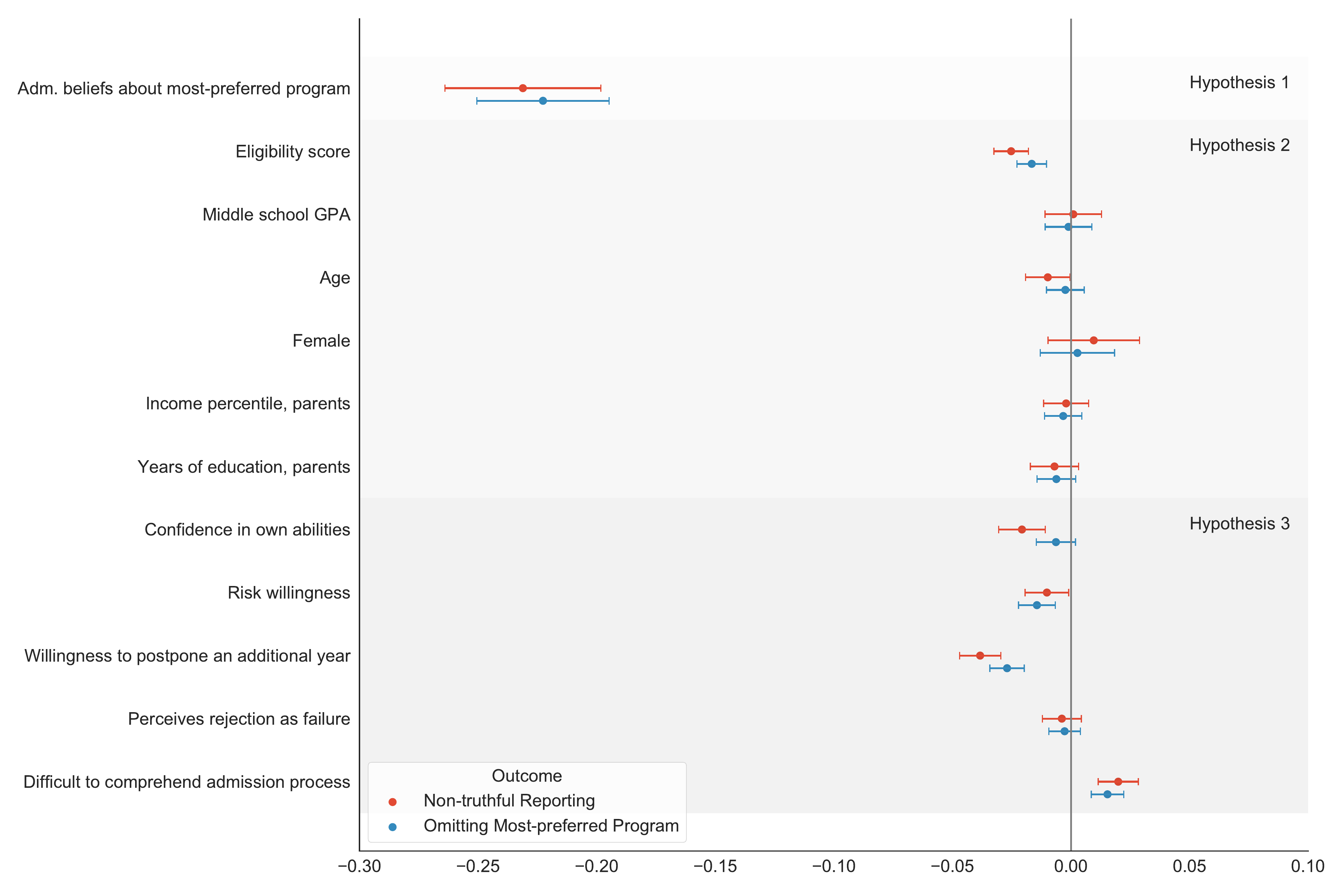}
    \end{subfigure}
\caption{Individual characteristics and truthful behavior; testing Hypotheses \ref{hypothesis:beliefs}, \ref{hypothesis:SES}, and \ref{hypothesis:Personality}. Including study program fixed effects.}
\label{fig:hyp_1_to_3_ols_fe}
\floatfoot{Notes:  The figure contains estimated coefficients from model (\ref{eq:model_hyp_1_2_3}) with 95\% confidence intervals. Compared to Figure~\ref{fig:hyp_1_to_3_ols_no_fe} the only difference is that this model is estimated with fixed effects at the study program level. The dependent variable is one of two measures of non-truthful behavior, see the legend. Each entry on the vertical axis corresponds to an explanatory variable and the horizontal axis measures the variable's coefficient size.
Standard errors of coefficients use robust estimation but are not clustered.}
\end{figure}

\begin{figure}[!ht]
    \begin{subfigure}[b]{.8\linewidth}
        \includegraphics[width=\linewidth]{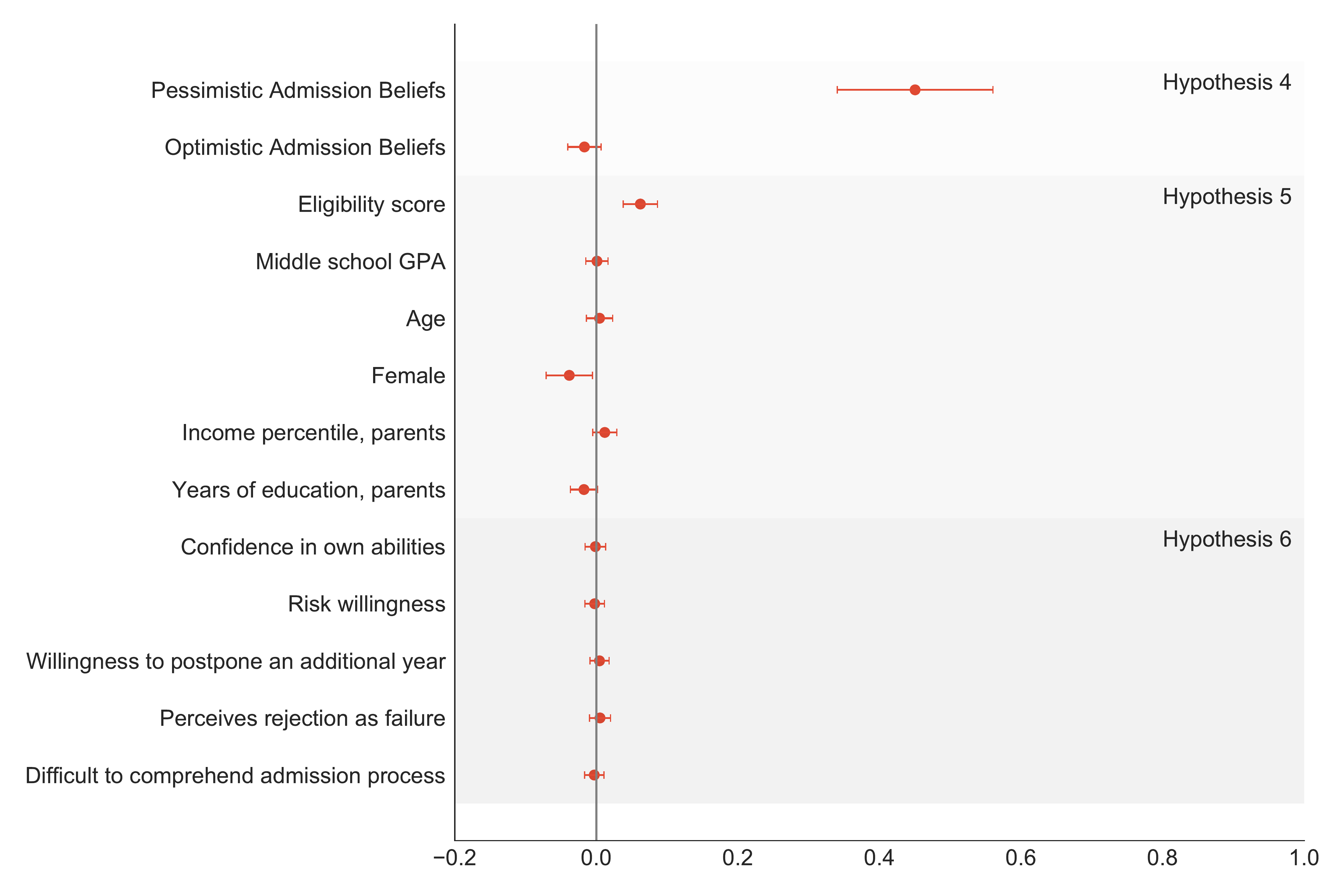}
    \end{subfigure}
\caption{Payoff-relevant Omission of Most-preferred Program. Including study program fixed effects.}
\label{fig:model_hyp_4_5_6_fe}
\floatfoot{Notes: The figure above shows coefficients from model (\ref{eq:model_hyp_4_5_6}) where payoff-relevant omission of most-preferred program is the outcome. Compared to Figure~\ref{fig:regret_skip} the only difference is that this model is estimated with fixed effects at the study program level. Coefficients are estimated with 95\% confidence intervals using robust standard errors.}
\end{figure}


\input{tables/linreg_robustness_2021_survey_no_fe}
\input{tables/clogit_res_offset_both_surveys_no_FE}

%% file: tables/admission_summary_stats.tex
\begin{table}[H]
\centering
\begin{tabular}{lrrr} \hline \hline
Year & Students & Academic Study Programs & Study Programs with Cutoff \\ \hline
2012 & 41,962    & 342            & 49\%                       \\ 
2013 & 44,490    & 341            & 51\%                       \\ 
2014 & 44,140    & 347            & 49\%                       \\ 
2015 & 45,246    & 344            & 52\%                       \\ 
2016 & 47,193    & 324            & 59\%                       \\ 
2017 & 45,916    & 325            & 61\%                       \\ 
2018 & 45,483    & 323            & 58\%                       \\ 
2019 & 44,533    & 323            & 56\%                       \\ 
2020 & 48,895    & 323            & 56\%                      \\ 
2021 & 48,796    & 327            & 45\%                      \\ \hline \hline
\end{tabular}
\caption{Summary Statistics for Danish Admission to Higher Education}
\label{tab:admission_summary_stats}
\floatfoot{Notes: We only consider students who apply for at least one academic study program.}
\end{table}

%% file: tables/linreg_robustness_2021_survey_no_fe.tex
\begin{table}[]
\centering
\small
\renewcommand{\arraystretch}{1.25} 
{
\def\sym#1{\ifmmode^{#1}\else\(^{#1}\)\fi}
\begin{tabular}{l*{4}{p{2cm}}}
\hline\hline
 & \multicolumn{2}{l}{\textbf{Reports}}  & \multicolumn{2}{l}{\textbf{Omits}} \\
  & \multicolumn{2}{l}{\textbf{non-truthfully}}  & \multicolumn{2}{l}{\textbf{most-preferred program}} \\
 & (1) & (2) & (3) & (4) \\  
\hline
\multicolumn{5}{l}{\textit{Hypothesis 1} } \\

Adm beliefs most-preferred program&      -0.272&            &      -0.270&            \\
            &     (0.019)&            &     (0.016)&            \\
Combined adm beliefs most-preferred program&            &     -0.326&            &     -0.328\\
           &            &     (0.022)&            &     (0.020)\\
\multicolumn{5}{l}{\textit{Hypothesis 2}} \\
Eligibility score&      -0.022&      -0.027&      -0.012&      -0.016\\
            &     (0.004)&     (0.004)&     (0.004)&     (0.003)\\
Middle school GPA&      -0.007&      -0.005&      -0.007&      -0.005\\
            &     (0.008)&     (0.008)&     (0.007)&     (0.007)\\
Age         &      -0.011&      -0.009&      -0.007&      -0.006\\
            &     (0.007)&     (0.007)&     (0.006)&     (0.005)\\
Female      &       0.032&       0.031&       0.004&       0.003\\
            &     (0.012)&     (0.012)&     (0.010)&     (0.010)\\
Income percentile, parents&      -0.005&      -0.005&      -0.010&      -0.010\\
            &     (0.007)&     (0.007)&     (0.006)&     (0.006)\\
Years of education, parents&       0.002&       0.002&      -0.000&       0.001\\
            &     (0.007)&     (0.007)&     (0.006)&     (0.006)\\
\multicolumn{5}{l}{\textit{Hypothesis 3}} \\
Confidence in own abilities&      -0.001&       0.001&       0.008&       0.011\\
            &     (0.007)&     (0.007)&     (0.006)&     (0.006)\\
Risk willingness&      -0.013&      -0.012&      -0.020&      -0.019\\
            &     (0.007)&     (0.006)&     (0.006)&     (0.005)\\
Willingness to postpone an additional year&      -0.036&      -0.034&      -0.027&      -0.025\\
            &     (0.006)&     (0.006)&     (0.005)&     (0.005)\\
Perceives rejection as failure&      -0.010&      -0.011&      -0.011&      -0.012\\
            &     (0.006)&     (0.006)&     (0.005)&     (0.005)\\
Difficult to comprehend admission process&       0.026&       0.023&       0.017&       0.015\\
            &     (0.006)&     (0.006)&     (0.005)&     (0.005)\\
\multicolumn{5}{l}{\textit{Hypothesis 8} } \\
Understands strategy-proofness &      -0.031&      -0.031&      -0.034&      -0.033\\
           &     (0.014)&     (0.014)&     (0.011)&     (0.011)\\

\hline
N           &    3,787        &    3,787      &    3,787       &    3,787         \\
Study program fixed effects&         No         &         No         &         No         &         No         \\
\hline\hline
\end{tabular}
}
\caption{Coefficient Estimates of Robustness Analyses}
\label{tab:robust_reg_2021}
\floatfoot{Notes: Outcome variable of equations (1) and (2) is non-truthful reporting. Outcome variable of equations (3) and (4) is omitting most-preferred program. Robust standard errors are reported in parenthesis. The models are estimated on survey respondents from the 2021 wave, since the 2020 wave did not include the required questions.}

\end{table}

%% file: tables/clogit_res_offset_both_surveys_no_FE.tex
\begin{table}[htbp]
    \centering
    \small
    \def\sym#1{\ifmmode^{#1}\else\(^{#1}\)\fi}
    \begin{tabular}{l*{3}{c}}
    \hline\hline
                &\multicolumn{1}{c}{(1)}&\multicolumn{1}{c}{(2)}&\multicolumn{1}{c}{(3)}\\
                &\multicolumn{1}{c}{Revealed Pref}&\multicolumn{1}{c}{Stated Pref}&\multicolumn{1}{c}{Stated Pref}\\
    \hline
                            &            &            &            \\
    Distance (1000 km)      &          -1&       -1&      -1\\
                            &           &      &      \\
    Program Peer Quality    &        9.32&        7.23&        9.49\\
                            &      (0.10)&      (0.08)&      (0.10)\\
    Same Gender Share       &        1.71&        1.60&        1.78\\
                            &      (0.04)&      (0.04)&      (0.04)\\
    Program Peer Parents Inc&        4.51&        3.22&        5.35\\
                            &      (0.19)&      (0.18)&      (0.16)\\
    \hline
    N                       &   6,888,528&   6,888,528&   5,973,743\\
    FE                      &          No&          No&          No\\
    Revealed Preferences    &         Yes&          No&          No\\
    Truthful Assumption     &          No&          Yes&         No\\
    Stability Assumption    &          No&          No&         Yes\\
    \hline\hline
    \end{tabular}
    \caption{Comparison of demand models (Without FE)}
    \label{tab:demand_est_offset_no_FE}
    \floatfoot{Notes: This table provides regression coefficients from estimating demand models under the three approaches of (1) revealed preferences, (2) baseline model, and (3) using the stability assumption, see text for details. The models are estimated with standard errors clustered at the study program level.}
    \end{table}